\documentclass[twocolumn,twocolappendix, trackchanges]{aastex63}
\hypersetup{linkcolor=cyan,citecolor=cyan,filecolor=green,urlcolor=blue}

\usepackage{listings}
\usepackage{soul}
\usepackage{enumitem}
\usepackage{lipsum}
\usepackage{threeparttable}
\usepackage{amsmath}

\received{December 8, 2020}
\revised{XX}
\accepted{XX}
\submitjournal{ApJ}

\shorttitle{Star-Gas Misalignment in Galaxies: II}
\shortauthors{D. J. Khim et al.}
\graphicspath{{./}}
\begin{document}

\title{Star-Gas Misalignment in Galaxies: II. Origins Found from the Horizon-AGN Simulation}

\correspondingauthor{Sukyoung K. Yi}
\email{E-mail: yi@yonsei.ac.kr}

%\author[0000-0002-0786-7307]{Greg J. Schwarz}
\author{Donghyeon J. Khim}
\affiliation{Department of Astronomy and Yonsei University Observatory, Yonsei University, Seoul 03722, Republic of Korea}
%\collaboration{1}{(AAS Journals Data Scientists collaboration)}
%\altaffiliation{AASTeX v6+ programmer}

\author{Sukyoung K. Yi}
\affiliation{Department of Astronomy and Yonsei University Observatory, Yonsei University, Seoul 03722, Republic of Korea}

\author{Christophe Pichon}
\affiliation{Institut d’ Astrophysique de Paris, Sorbonne Universités, et CNRS, UMP 7095, 98 bis bd Arago, 75014 Paris, France}
\affiliation{Korea Institute of Advanced Studies (KIAS), 85 Hoegiro, Dongdaemun-gu, Seoul, 02455, Republic of Korea}

\author{Yohan Dubois}
\affiliation{Institut d’ Astrophysique de Paris, Sorbonne Universités, et CNRS, UMP 7095, 98 bis bd Arago, 75014 Paris, France}

\author{Julien Devriendt}
\affiliation{Dept of Physics, University of Oxford, Keble Road, Oxford OX1 3RH, UK}

\author{Hoseung Choi}
\affiliation{Korea Astronomy and Space Science Institute 776, Daedeokdae-ro, Yuseong-gu, Daejeon, Republic of Korea}

\author{Julia J. Bryant}
\affiliation{Sydney Institute for Astronomy (SIfA), School of Physics, The University of Sydney, NSW 2006, Australia}
\affiliation{Australian Astronomical Optics, AAO-USydney, School of Physics, University of Sydney, NSW 2006, Australia}
\affiliation{ARC Centre of Excellence for All Sky Astrophysics in 3 Dimensions (ASTRO 3D)}

\author{Scott M. Croom}
\affiliation{Sydney Institute for Astronomy (SIfA), School of Physics, The University of Sydney, NSW 2006, Australia}
\affiliation{ARC Centre of Excellence for All Sky Astrophysics in 3 Dimensions (ASTRO 3D)}

%% Note that the \and command from previous versions of AASTeX is now
%% depreciated in this version as it is no longer necessary. AASTeX 
%% automatically takes care of all commas and "and"s between authors names.

%% AASTeX 6.3 has the new \collaboration and \nocollaboration commands to
%% provide the collaboration status of a group of authors. These commands 
%% can be used either before or after the list of corresponding authors. The
%% argument for \collaboration is the collaboration identifier. Authors are
%% encouraged to surround collaboration identifiers with ()s. The 
%% \nocollaboration command takes no argument and exists to indicate that
%% the nearby authors are not part of surrounding collaborations.

%% Mark off the abstract in the ``abstract'' environment. 

\begin{abstract}

There have been many studies aiming to reveal the origins of the star-gas misalignment found in galaxies, but there still is a lack of understanding of the contribution from each formation channel candidate. We aim to answer the question by investigating the misaligned galaxies in Horizon-AGN, a cosmological large-volume simulation of galaxy formation.
There are 27,903 galaxies of stellar mass $M_* > 10^{10} M_\odot$ in our sample, of which 5,984 are in a group of the halo mass of $M_{200} > 10^{12} M_\odot$. 
We have identified four main formation channels of misalignment and quantified their level of contribution: mergers (35\%), interaction with nearby galaxies (23\%), interaction with dense environments or their central galaxies (21\%), and secular evolution including smooth accretion from neighboring filaments (21\%).
We found in the simulation that the gas, rather than stars, is typically more vulnerable to dynamical disturbances; hence, misalignment formation is mainly due to the change in the rotational axis of the gas rather than stars, regardless of the origin. We have also inspected the lifetime (duration) of the misalignment. The decay timescale of the misalignment shows a strong anti-correlation with the kinematic morphology ($V/{\sigma}$) and the cold gas fraction of the galaxy. The misalignment has a longer lifetime in denser regions, which is linked with the environmental impact on the host galaxy. There is a substantial difference in the length of the misalignment lifetime depending on the origin, and it can be explained by the magnitude of the initial position angle offset and the physical properties of the galaxies.

%word limit : 250

\end{abstract}

%% Keywords should appear after the \end{abstract} command. 
%% See the online documentation for the full list of available subject
%% keywords and the rules for their use.

%galaxies: structure

%% From the front matter, we move on to the body of the paper.
%% Sections are demarcated by \section and \subsection, respectively.
%% Observe the use of the LaTeX \label
%% command after the \subsection to give a symbolic KEY to the
%% subsection for cross-referencing in a \ref command.
%% You can use LaTeX's \ref and \label commands to keep track of
%% cross-references to sections, equations, tables, and figures.
%% That way, if you change the order of any elements, LaTeX will
%% automatically renumber them.
%%
%% We recommend that authors also use the natbib \citep
%% and \citet commands to identify citations.  The citations are
%% tied to the reference list via symbolic KEYs. The KEY corresponds
%% to the KEY in the \bibitem in the reference list below. 

%%%%%%%%%%%%%%%%%%%%%%%%%%%%%%%%%%%%%%%%%%%%%%%%%%

%%%%%%%%%%%%%%%%% BODY OF PAPER %%%%%%%%%%%%%%%%%%

\section{Introduction}

As stars form from gas, stars and gas of a galaxy are naturally expected to share the same or similar spin orientation, assuming the overall angular momentum is conserved. However, spectral observations have found that some galaxies have different axes of rotation between stars and gases \citep[e.g.,][]{1975PASP...87..965U, 1992ApJ...401L..79B, 1992ApJ...394L...9R,1996MNRAS.283..543K, 2001AJ....121..140K, 2016MNRAS.455.2508S}. Moreover, recent integral field spectroscopy (IFS) observations confirmed the presence of misaligned galaxies regardless of the shape and environment \citep[e.g.,][]{2006MNRAS.366.1151S, 2011MNRAS.412L.113C, 2015A&A...581A..65C, Davis+11, Serra+14, 2014A&A...568A..70B, 2015A&A...582A..21B, 2015MNRAS.452....2K, 2016MNRAS.461.2068K, Jin+16, 2019MNRAS.483..458B}. In particular, \cite{2019MNRAS.483..458B}, using the Sydney-AAO Multi-object Integral field spectrograph (SAMI) Galaxy Survey \citep{2015MNRAS.447.2857B, 2012MNRAS.421..872C}, reported that about 11\% of observed galaxies have different rotation axes between stars and gases by more than 30\,degrees.

A number of observational and theoretical (simulation) studies have been conducted to find the origin of the star-gas misaligned galaxies. Through various previous studies in both fields, (i) galaxy merger \citep[e.g.,][]{1990ApJ...361..381B, 1991Natur.354..210H, 1996ApJ...471..115B, 1998ApJ...499..635B, 2001Ap&SS.276..909P, 2009MNRAS.393.1255C, 2006MNRAS.366..182B}, (ii) continuous accretion of cold gas through the cosmic filaments \citep[e.g.,][]{1996ApJ...461...55T, 2003A&A...401..817B, 2008ApJ...689..678B, 2013MNRAS.428.1055A, 2014MNRAS.437.3596A, 2015MNRAS.451.3269V}, and (iii) interactions with nearby galaxies \citep[e.g.,][]{2004A&A...426...53D, 2006MNRAS.370.1565C} have been suggested as the possible origins of the misalignment. 
Many of these studies, however, used only a small number of galaxies that are often generated under ``idealized'' assumptions.
Such studies hence lack statistical significance, especially considering that the star-gas misalignment is a highly nonlinear phenomenon as a result of various competing effects.
In particular, if there are multiple origins for misalignments, it is critical to have a large number of galaxies in the analysis in order to quantify the level of contribution from those channels.
For example, there is still controversy over whether galactic mergers are the main source of the misalignment. While the various studies described above showed that misalignment might occur through galactic mergers, \cite{2019MNRAS.483..458B} and \cite{2019ApJ...878..143S} reported that mergers do not significantly contribute to the formation of misalignment.

The large-volume simulations that have lately been made available (Horizon-AGN; \citealp{2014MNRAS.444.1453D}, Eagle; \citealp{Eagle}, and Illustris; \citealp{Illustris}) provide a large number of galaxies of various properties in various environments and thus solve the problem. 
Such simulations have a large volume of up to (100\,Mpc/h$)^3$ that contains more than 100,000 galaxies.
We can also inspect the time evolution of specific galaxies, which enables us to simultaneously identify and verify the processes behind the misalignment.
Besides, it allows us to study the different processes that are more important in different environments.
This includes not just the various halo sizes of galaxies but also the large-scale effects such as cold gas flows from filaments.
In summary, large-volume simulations allow us to inspect the effects of various processes in a comprehensive manner.
We, therefore, choose one of the recent large-volume simulations, Horizon-AGN, to study the origin of misalignment in our investigation.

In the previous paper \citep[][Paper I]{Khim+}, we investigated the properties of misaligned galaxies using the Horizon-AGN simulation and the observational data from SAMI \citep{2019MNRAS.483..458B}. We hereby summarize the main results of Paper I. 
The earlier the morphological type of a galaxy, the higher the probability of misalignment and the greater the misalignment angle. More massive galaxies tend to show the misalignment more often, but this is likely a reflection of the morphology-misalignment relation given above and the mass-morphology relation \citep[e.g.,][]{2006MNRAS.373.1389C, 2010ApJ...709..644I, 2010ApJ...719.1969B, 2016MNRAS.463.3948D}. 
Gas-poor galaxies in Horizon-AGN tend to show the misalignment more often. This seemed more significant than expected from the widely-known relation between the morphology and gas fraction. 
However, a discrepancy is noted between Horizon-AGN and SAMI. 
Horizon-AGN showed a significantly higher misalignment fraction in denser environments than reported by SAMI and other IFS studies \citep[e.g.,][]{Davis+11, Jin+16}. 

Similar studies have also been performed recently based on different cosmological hydrodynamic simulations. \cite{2019ApJ...878..143S} investigated the properties of the counter-rotation in low mass galaxies using the Illustris simulation. \citet{Duckworth+20} also examined the misaligned galaxies by comparing the MaNGA observations and the IllustrisTNG100 simulation. Although these three studies, including Paper I, have used different simulations, they have found consistent results in the properties of the misaligned galaxies: (i) the misaligned galaxies tend to have a lower gas fraction than the aligned galaxies, (ii) there is a strong correlation between the misalignment fraction and the morphology of the galaxy.

In addition, \cite{2019ApJ...878..143S} investigated the origin and duration time of the misalignment. They reported that supermassive black hole feedback and gas stripping during fly-by passes through group environments were the two main channels of misalignment. They also reported that some galaxies maintained misaligned components for more than 2\,Gyr. We will show in the following sections that our study confirms some of their key results and presents additional results based on a different simulation.

In the present study, we investigate the origin and evolution of misalignment based on the Horizon-AGN simulation data. 
First, we scrutinize the change of rotational axes of stars and gases as a galaxy experiences a mass change in stars or gas due to various events including galactic interactions (Section~\ref{sec:massmass}). 
After that, we classify various formation channels for the star-gas misalignment and quantify the contribution of each channel (Section~\ref{sec:origin}). Next, we examine the environmental and physical properties of misaligned galaxies depending on their classified origins (Section~\ref{sec:property}). We finally investigate the lifetime of the misalignment, depending on the galaxy's physical properties, environments, and its misalignment origins (Section~\ref{sec:lifetime}).

%%%%%%%%%%%%%%%%%%%%%%%%%%%  Section 2  %%%%%%%%%%%%%%%%%%%%%%%%
\section{Methodology}
\label{sec:2}
\subsection{Horizon-AGN simulation}
\label{sec:hagn} % used for referring to this section from elsewhere

Horizon-AGN \citep{2014MNRAS.444.1453D}, one of the major recent collaborative efforts of cosmological large-volume hydrodynamic simulations, was performed using the adaptive mesh refinement code {\sc{ramses}} \citep{2002A&A...385..337T}, based on the cosmological context from the seven-year Wilkinson Microwave Anisotropy Probe results \citep{2011ApJS..192...18K}. The simulation includes gas cooling, star formation, and feedback from stars and supermassive black holes.
The volume of the simulation box is (100\,Mpc\,$h^{-1}$)$^{3}$, and the maximum (smallest) force resolution is about 1\,kpc. The mass resolution is $8\times10^7 M_{\sun}$ for dark matter and $2\times10^6 M_{\sun}$ for stellar particles. A total of 787 snapshots are in Horizon-AGN, each with a time interval of about 17\,Myr. However, only 61 snapshots with a time interval of about 250\,Myr contain information on gas, dark matter, and sink particles. As our research requires both stars and gas, from now on, the word {\em snapshot} refers only to these 61 ``gas snapshots.'' Meanwhile, the all 787 snapshots were used to analyze the merger trees, merger processes (Section~\ref{sec:hagn_merger}), the path of galaxy interaction (Section~\ref{sec:origin}), and perturbation index (Section~\ref{sec:origin_non_merger}). For more details on the simulation, readers are referred to \cite{2014MNRAS.444.1453D}.

\subsection{Galaxy and group identification}
\label{sec:hagn_gal} % used for referring to this section from elsewhere

Horizon-AGN galaxies were identified by using HaloMaker through the AdaptaHOP algorithm \citep{2004MNRAS.352..376A}, with the most massive sub-node mode \citep{2009A&A...506..647T} applied for stellar particles. A group of stars had to contain at least 50 star particles to be classified as galaxies, which corresponds to $1.7\times10^8 M_{\sun}$. Using this method, 124,744 galaxies were identified at $z=0.018$, the last gas snapshot of Horizon-AGN.

As mentioned in Paper I, galaxies with a small number of stellar particles are not suitable for studying the kinematics of galaxies \citep[e.g.,][]{2015MNRAS.454.1886S, 2016MNRAS.463.3948D, 2017MNRAS.467.3083R, 2017MNRAS.468.3883P}. 
Therefore, we used only massive galaxies with a stellar mass of more than $10^{10}M_{\sun}$, corresponding to about 3,000 stellar particles; the same criterion used in the preceding study. The number of galaxies above the criterion is 27,903 out of 124,744 at $z=0.018$. 

The stellar $V/{\sigma}$ represents the kinematic property of a galaxy, where $V$ and ${\sigma}$ are the rotational velocity and the velocity dispersion of the galaxy, respectively. The galaxies with a higher $V/{\sigma}$ tend to have disk-shaped structures whereas the galaxies with a lower $V/{\sigma}$ tend to have spheroidal structure. We measured $V/{\sigma}$ as done by \cite{2016MNRAS.463.3948D}. After that, we classified the galaxies into early-type galaxies (ETG) and late-type galaxies (LTG), based on the $V/{\sigma}$ cut of 1, as we performed in Paper I \citep[see also,][]{2016MNRAS.463.3948D}. Note that we have already analyzed the misaligned galaxies with respect to the $V/{\sigma}$ ratio in Paper I. 

Galaxy groups were defined as follows, using the same method as in Paper I \citep[see,][]{Khim+}. We first identified dark matter halos with $M_{\rm vir} \geq 10^{12}M_{\sun}$ and defined a galaxy located within its 1.5 virial radii ($R_{\rm 200}$) as a member. If a dark matter halo had more than three member galaxies ($M_* > 10^{10}M_{\sun}$), it was defined as a galaxy group. Based on the above definition, we identified a total of 606 galaxy groups and a total of 5,984 member galaxies in the group environment at $z=0.018$. The most massive group had 208 group members and a halo mass of $7.4\times10^{14}M_{\sun}$ (14.87 in logarithm solar-mass). This definition is the same as that of Paper I, but we did not separate ``clusters'' from groups this time.

\subsection{Galactic gas and gas detection limit}
\label{sec:hagn_gas}

We use a linear cut in the logarithmic density-temperature plane using Equation~(\ref{eq:torrey}) from \cite{2012MNRAS.427.2224T} to separate the gases in the simulation into galactic cold gas and non-cold surroundings:
\begin{equation}
    \log(T/[K])=6+0.25\log(\rho/10^{10} [M_\odot h^2 \rm{kpc} ^{-3}]).
	\label{eq:torrey}
\end{equation}
``Galactic cold gas,'' the gas below the density-temperature criterion, has a temperature of about 10,000 -- 30,000 K, depending on the density. The cold gas is used for star formation in the simulation and represents the kinematic property of the interstellar medium. Also, they have similar properties to the gas component observed by H$\alpha$ emission lines ($\simeq 10,000$ K) in IFS, such as SAMI. However, it should be noted that Horizon-AGN cannot resolve extremely cold gas, such as molecular gas. Contrary to the galactic cold gas, (non-cold) ``surroundings,'' the high-temperature side, corresponds to the intergalactic gas or intra-group medium. This relatively hot gas is used to measure ram pressure \citep{1972ApJ...176....1G, 2000Sci...288.1617Q} in Section~\ref{sec:property_env}.

Galaxies that do not have sufficient gas need to be excluded from the sample because their gas rotation cannot be measured or will show a large error if measured. In our previous paper \citep{Khim+} comparing SAMI and Horizon-AGN, it was assumed that the gas flux detection limit would be linked with the cold gas fraction, the mass ratio of cold gas to stars ($M_{\rm cold\,gas}/M_{\rm *}$). We measured the gas fraction within one effective radius ($R_{\rm eff}$), encompassing half of the total stellar mass (half projected stellar mass) of the galaxy, which is similar to the range observed by SAMI. We compared the misalignment fractions in SAMI and Horizon-AGN, and suggested that galaxies with the gas fraction higher than 3\% were suitable for use as samples. With this criterion, 26,222 out of 27,903 galaxies at $z=0.018$ have enough gas fraction and are considered observable.

\begin{figure*}
	% To include a figure from a file named example.*
	% Allowable file formats are eps or ps if compiling using latex
	% or pdf, png, jpg if compiling using pdflatex
	\includegraphics[width=\linewidth]{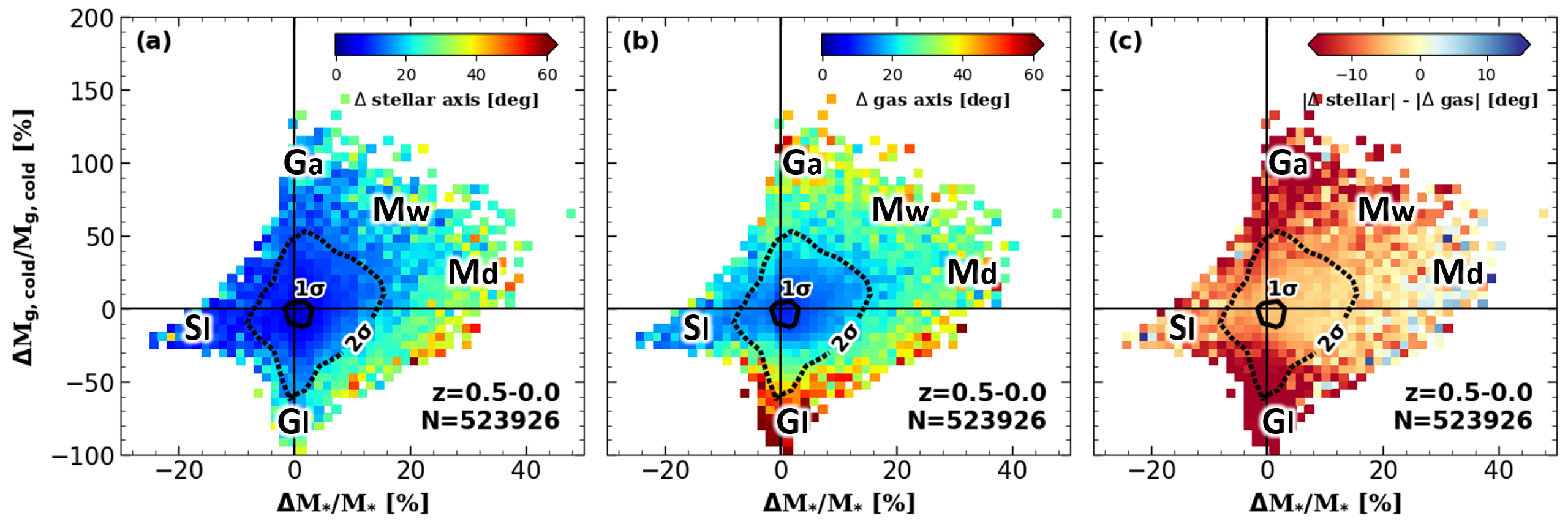}
    \caption{The rotational axis changes depending on the mass changes of stars and cold gas since $z=0.5$. The mass changes of stars (X-axis) and gas (Y-axis) during one gas snapshot interval ($\simeq$ 0.25\,Gyr) is expressed in percentages. The black contours show the $1 {\sigma}$ (solid line) and $2 {\sigma}$ (dotted line) distributions of galaxies. To ensure statistical significance, each pixel contains at least 10 galaxies. The left two panels show the changes in the rotational axis for the stars (Panel (a)) and cold gas (Panel (b)). Panel (c) shows the differences between the two rotational axes. We have also marked a few notable galactic phenomena in the figure: ``Ga'' (gas accretion), ``Mw'' (wet merger), ``Md'' (dry merger), ``Gl'' (gas mass loss), and ``Sl'' (stellar mass loss.)
    The changes in the rotational axis seem to be linked with the changes in the stellar/gas mass of a galaxy. Almost all regions in Panel (c) are red, indicating that the rotational axis is more easily changed for the gases than for the stars.}
    \label{fig:image_massmass}
\end{figure*}

\subsection{Rotational axes and star-gas misalignment}

In order to calculate the rotational axes of a galaxy, we measured the angular momentum of stellar particles and cold gas cells with respect to the galactic center. Then, the direction of the net angular momentum of each component was regarded as the rotational axis. The position angle (PA) offset, which is the difference between the rotation axes of stars and gases (misaligned angle), has a range from 0 (co-rotation or aligned) to 180\,degrees (counter-rotation). We calculated the stellar and gas rotational axes inside $1R_{\rm eff}$ of the galaxy, similarly to that of many IFS observations. The galaxies with PA offset larger than 30\,degrees were classified as misaligned galaxies to be consistent with SAMI and other observational studies. Among the galaxies with sufficient gas contents, 2,662 galaxies (10.2\%) were misaligned and used as the main targets in this study. Although the snapshots used in this study are slightly different from those used in Paper I ($z=0.055$), the misalignment features in terms of number fraction, properties, and PA offset distribution of misaligned galaxies are almost identical.

\subsection{Galaxy mergers}
\label{sec:hagn_merger}

We have built merger trees to link all the identified galaxies reaching far below our mass cut in all 787 snapshots. After merger trees are made, we have investigated the stellar particle exchange between interacting galaxies as follows. The ``initial masses'' of galaxies are measured at the time the secondary (less-massive) galaxy passes through the virial radius of the primary (more-massive) galaxy. The merger was defined when more than 50\% of the initial mass of the secondary galaxy is consumed by the primary galaxy. We confirm that a different choice of the cut would not cause a notable difference as long as it is in a reasonable range (50 -- 90 \%).

We used a narrow definition for the duration of mergers to distinguish mergers and interactions as much as possible. The beginning of a merger is defined as the moment when the exchange of star particles begins. Similarly, the endpoint of the merger is defined as the moment when the exchange stops, or when the secondary galaxy is no longer detected (coalescence).

We define three classes of mergers based on the mass ratio: (1) major merger when $M_{\rm *,s}/M_{\rm *,p} \geq 1/4$, (2) minor merger when $ 1/4 > M_{\rm *,s}/M_{\rm *,p} \geq 1/50$, and (3) tiny merger when $M_{\rm *,s}/M_{\rm *,p} < 1/50$, where $M_{\rm *,p}$ and $M_{\rm *,s}$ denote the stellar masses of the primary and secondary galaxies, respectively. As described earlier in Section~\ref{sec:hagn_gal}, our galaxy finding scheme detects all the galaxies above $1.7\times10^8 M_{\sun}$, while we use the galaxies above $10^{10} M_{\sun}$ as main targets. Therefore, the galaxies near the mass limit of $10^{10} M_{\sun}$ would have only major and minor mergers, but few tiny mergers. As a result, a total of 21,122 major mergers and 50,848 minor mergers have taken place in Horizon-AGN since $z=3$.

%%%%%%%%%%%%%%%%%%%%%%%%%%%  Section: Results  %%%%%%%%%%%%%%%%%%%%%%%%
\section{Rotational axis changes}
\label{sec:massmass}

The most obvious candidate for the origin of misalignment is interactions between galaxies. Galaxy interactions likely cause changes in the stellar and gaseous masses of galaxies \citep[e.g.,][]{2014MNRAS.445L..46W}; and thus, we first inspect the mass changes of the aligned and misaligned galaxies in the Horizon-AGN sample.

Figure~\ref{fig:image_massmass} shows the rotational axis changes depending on the mass changes of stars and cold gas. We first select all galaxies at $z \leq 0.5$ containing sufficient stellar ($M_* > 10^{10}M_{\sun}$) and cold gas mass ($M_g > 10^{9}M_{\sun}$). After that, the mass changes of stars (X-axis) and gas (Y-axis) are measured during one gas snapshot interval ($\simeq$ 0.25\,Gyr) and expressed in percentages. We measure the rotational axis changes for the stars (Figure~\ref{fig:image_massmass}-(a)) and the gas (Figure~\ref{fig:image_massmass}-(b)) with pixels color-coded by their mean changes in the rotational axis. The redder the color on these two panels, the larger the mean value of the rotational axis change. Each pixel contains at least 10 galaxies to ensure statistical significance. The black contours show the $1 {\sigma}$ (solid line) and $2 {\sigma}$ (dotted line) distributions of galaxies. 

We mark the regions for a few notable galactic phenomena in Figure~\ref{fig:image_massmass}. 
First, the galaxies that receive gas from cosmic filaments or nearby galaxies are marked in Region ``Ga'' (gas accretion). When galaxies go through a merger, they would get a significant amount of stars and gas, which are marked in Regions ``Mw'' (wet merger) and ``Md'' (dry merger). Region ``Gl'' (gas mass loss) marks the galaxies with gas mass loss due to strong outflow or ram pressure stripping, and Region ``Sl'' (stellar mass loss) marks the galaxies with stellar mass loss such as tidal stripping.

However, it should be noted that a galaxy may be classified as more than one type in this diagram throughout a full event (longer than 0.25\,Gyr). For example, a major merger may first show a dramatic enhancement in gas mass hence being classified as Region Ga, then later appear in Region Mw by accreting a substantial amount of stars, and finally appear in Region Gl because of gas loss due to the star formation itself and to the stellar feedback from enhanced star formation. In this sense, the regional classification is not uniquely given to a case.

\subsection{Stellar axis}

The stellar rotational axis is influenced when there is a significant accretion of stellar mass (Regions Mw, Md), as shown in the Figure~\ref{fig:image_massmass}-(a). 
In contrast, the change in gas mass regardless of inflow or outflow (Regions Ga, Gl) does not influence the stellar rotational axis significantly. 
Some of our target galaxies show decrease in the stellar mass (Region Sl) mostly due to galaxy interactions, but it hardly affects the stellar rotational axis, either.

\subsection{Gas axis}

The gas rotational axis changes are shown in Figure~\ref{fig:image_massmass}-(b).
The gas rotational axis is influenced by an inflow of gas from outside (Regions Ga, Mw), which is easily expected. We found that it is also influenced by an accretion of stars (Region Md), which is mostly due to interactions.
Interestingly, the greatest magnitude of the gas rotational axis change is found in the galaxies that experience a significant loss of gas while maintaining their stellar mass (Region Gl).
Such a major change in the rotational axis is not visible in any other region in Panels (a) or (b).
We have already demonstrated in Paper I that misalignment can occur when a galaxy loses its gas while falling into the central region of clusters, which has been confirmed by the study based on the Illustris simulation \citep{2019ApJ...878..143S}.
We found a considerable number of galaxies within Region Gl in field environments as well. Most of these galaxies appear to be experiencing a gas loss due to the active galactic nuclei (AGN) or stellar feedback. 
These massive gas loss events seemed to be linked with mergers.

\subsection{Different motions of two axes}
Figure~\ref{fig:image_massmass}-(c) shows the difference between the rotational axis changes of stars and gas: that is, Panel (c) is Panel (a) minus Panel (b), pixel by pixel. 
Blue pixels mark the galaxies that show larger axis changes in stars than in gas, whereas red pixels show larger changes in gas than in stars. 
The difference away from zero (red or blue) means that the PA offset, the angle between the two axes, has changed. Thus, these colors are closely linked with star-gas misalignment. 
Almost all regions in Panel (c) are red, indicating that the rotational axis is more easily changed for the gases than for the stars. This means that the main drivers of misalignment are the processes that influence gases rather than stars.
They are for example gas inflow (Regions Ga, Mw) and outflow (Region Gl).

Regions Mw, Md and Sl show the galaxies that experienced stellar mass changes more dramatically than other regions. These regions too show some misalignment (red or blue pixels); yet, the degree of misalignment is lower than Regions Ga and Gl, where stellar masses hardly changed. This means that galaxy interactions that involve stellar mass changes may not be the primary driver of misalignment, unlike our expectation at the beginning of this section.

\section{Origin of star-gas misalignment}
\label{sec:origin}

Let us investigate the origin of the misalignment in the Horizon-AGN galaxies in the local Universe based on the clues from Section~\ref{sec:massmass}. We classify the ``merger-driven'' misaligned galaxies using the time difference between merger events and misalignment epoch (Section~\ref{sec:origin_merger}). The merger-driven misaligned galaxies are subclassified depending on their merging mass ratio (e.g., major merger, minor merger). On the other hand, the non-merger-driven misaligned galaxies are subdivided into ``group-driven,'' ``interaction-driven,'' and ``secularly-driven,'' based on their environment or the presence of interactions between nearby galaxies (Section~\ref{sec:origin_non_merger}). 

\subsection{Merger-driven misalignment}
\label{sec:origin_merger} % used for referring to this section from elsewhere

As seen in the Figure~\ref{fig:image_massmass}, galaxies undergoing mergers have a large accretion of stars and gases, thus the merging galaxy can easily change their rotational axes. In this section, we will examine the process of merger-driven misalignment and quantify the level of contribution to misalignment formation.

\subsubsection{Mergers and misalignment}
\label{Mergers and misalignment}

\begin{figure}
	% To include a figure from a file named example.*
	% Allowable file formats are eps or ps if compiling using latex
	% or pdf, png, jpg if compiling using pdflatex
	\includegraphics[width=\columnwidth]{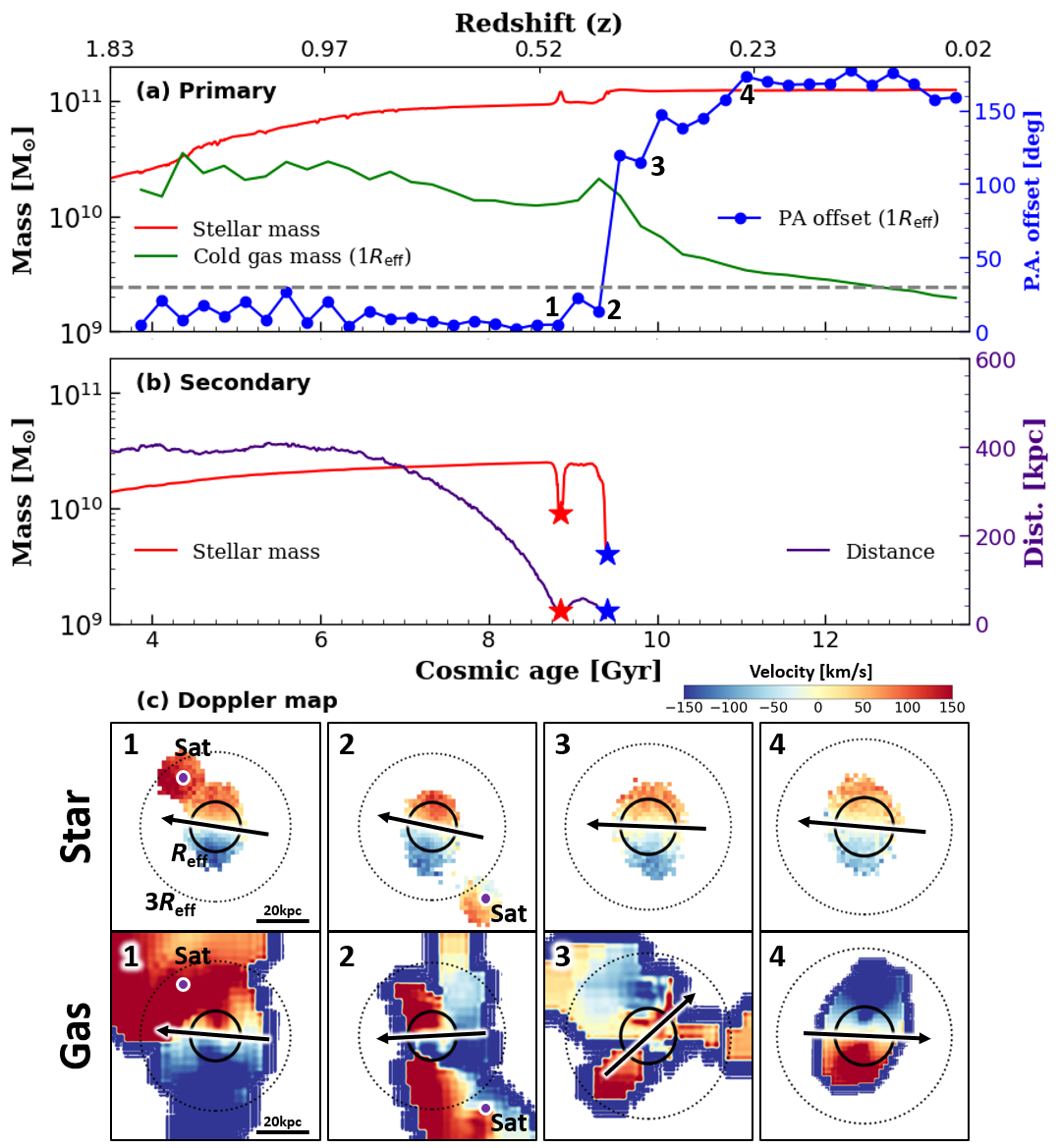}
    \caption{An example of a merger-driven misalignment formation with a stellar mass ratio about 1:4. We number particular snapshots along the trajectory (Stages 1 -- 4). Panel (a): the stellar mass (red), cold gas mass (green), and PA offset (blue) of the primary galaxy. The horizontal dashed line shows $30\,^{\circ}$, which is the criterion for the misalignment. Panel (b): the stellar mass (red) of the secondary galaxy and the distance (violet) from the primary galaxy. The red and blue star marks show the first and second pericenters, respectively. Panel (c): stellar and cold gas Doppler maps corresponding to each stage and the gas rotational axis at $1\,R_{\rm eff}$. The effective radius and $3\,R_{\rm eff}$ are expressed as solid and dotted lines, respectively. If the secondary galaxy is within the field of view, its center is marked with a purple dot (Stages 1--2).}
    \label{fig:image_seq_merger}
\end{figure}

Figure~\ref{fig:image_seq_merger} shows an example of merger-driven misaligned galaxies. Panel (a) shows the stellar mass (red), gas mass (green), and PA offset (blue) of the primary (target) galaxy as a function of time. The criterion for the misalignment, 30\,degrees, is expressed with a gray horizontal dashed line. Some snapshots are numbered (Stages 1--4) to illustrate the merger-driven misalignment formation process. In Panel (b), the stellar mass of the secondary galaxy (red) and the distance between the two galaxies (violet) are presented. We mark the first (the red star mark) and the second (the blue star mark) pericenter passes in the panel. In the case where the merging galaxies are very close to each other, HaloMaker cannot accurately distinguish the two galaxies. Thus, the stellar mass curve of the secondary galaxy in particular can be distorted or even partially disconnected near the pericenter. In Panel (c), projected stellar and cold gas Doppler maps corresponding to each step is shown. The effective radius and $3\,R_{\rm eff}$ are expressed in solid and dotted circles, respectively. Black arrows show the rotational axes within $1\,R_{\rm eff}$. When the secondary galaxy enters within the field of view, the center of the secondary galaxy is expressed with a purple dot (Stages 1--2). Coalescence happened near Stage 2.

\begin{figure}
	% To include a figure from a file named example.*
	% Allowable file formats are eps or ps if compiling using latex
	% or pdf, png, jpg if compiling using pdflatex
	\includegraphics[width=\columnwidth]{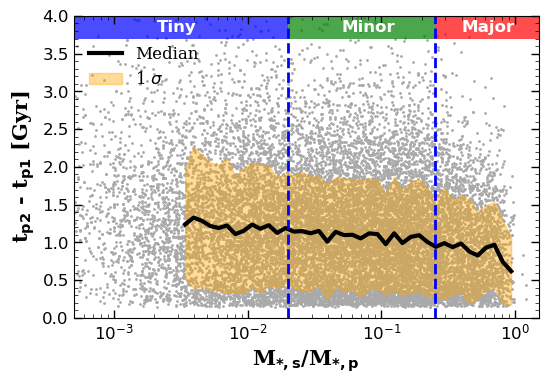}
    \caption{The orbiting period of the Horizon-AGN mergers as a function of their stellar mass ratio. We define the orbiting period as the time difference between the first and second pericenters. The merger events since $z=1$ are expressed using gray dots. The black curve shows the median value, and the shaded region shows a ${1 \sigma}$ error. Blue dashed lines show demarcations between major, minor, and tiny mergers.}
    \label{fig:image_merger_time}
\end{figure}

The secondary galaxy infalls into the primary galaxy with the aligned orbit with respect to the pre-existing stellar rotational axis. Also, the stellar mass ratio between the two galaxies is about 1:4. Thus, during Stages 1--4, the merger does not affect the stellar rotational axis significantly.

The gas axis, on the other hand, shows more dramatic features.
Before the merger, the primary galaxy has a well-developed aligned gas disk (Stage 1). However, as the secondary galaxy passes through the pericenter for the first time (red star), the cold gas of the secondary galaxy gradually begins to flow to the primary galaxy, affecting the PA offset of the primary galaxy (Stage 2). At the second pericenter pass, a large amount of cold gas, nearly as much as the pre-existing gas, flows into the $1\,R_{\rm eff}$ region, generating a strong star-gas misalignment. 
At this stage, the gas rotation of the primary galaxy is unstable due to the massive gas outflow (Stage 3). Finally, the gas disk stabilizes with a misaligned rotational axis (Stage 4).

We describe the trajectory of this case in the format of Figure~\ref{fig:image_massmass}.
The main galaxy is a typical galaxy with the aligned axes between stars and gas near the origin in the panel (i.e., coordinates 0,0). 
At Stage 1, the galaxy gets a significant amount of cold gas and moves to Region Ga in Figure~\ref{fig:image_massmass}. After that, the galaxy moves to the right during the merger process (Stage 2) according to its merging mass ratio. The galaxy gives off its cold gas and moves downward (Stage 3).
After the merger, the galaxy eventually returns to the origin in the panel and remains misaligned.

This and many other galaxies show a clear correspondence between the formation of misalignment and the gas mass change, which motivated us to focus on the gas and stellar mass changes during the merging and interacting events. However, we also note the possibility of position angle change through tidal effects between interacting galaxies even if there is little exchange of stars or gas.

\subsubsection{Merger-driven misalignment fraction}
\label{sec:origin_merger_frac}
The epoch of a merger event and the emergence of misalignment do not necessarily coincide. Some galaxies show misalignment even before the merging process becomes violent, for example, through gas accretion. 
Other galaxies show misalignment as a result of a merger, for example, through stellar accretion. These changes are thought to occur within a merging timescale frame.

The merging timescale is considered to be a couple of times the crossing time \citep[e.g.,][]{1978MNRAS.184..185W, 1981MNRAS.197..179G, 1988ApJ...331..699B}. It is however difficult to imagine a galaxy that is just touching the virial radius of another (target) galaxy affecting its rotational axes of gas and stars. It is more probable that a shorter distance is needed. An easy and practical way of estimating the crossing time is to measure the time difference between the first and the second pericenter passes, which is shown in Figure~\ref{fig:image_merger_time}. We here show all the mergers since $z=1$ for which we are able to identify both the first and second pericenter passes. While the data appear to be wildly scattered in the diagram, the median as a function of the galaxy mass ratio is steady and well defined. With a slow decay with increasing mass ratio, the typical value is roughly between 1.0 and 1.5\,Gyr. Therefore, the causality connection between merger and misalignment can be searched roughly within this time frame.

We now examine the correlation between the two events: mergers and misalignment formation. We use the merger trees to identify when the galaxies undergo mergers and compare those with the epochs when the galaxies develop the misalignment. In the case when the merger takes place across multiple snapshots, we select the nearest snapshot to the emergence of the misalignment.

\begin{figure}
	% To include a figure from a file named example.*
	% Allowable file formats are eps or ps if compiling using latex
	% or pdf, png, jpg if compiling using pdflatex
	\includegraphics[width=\columnwidth]{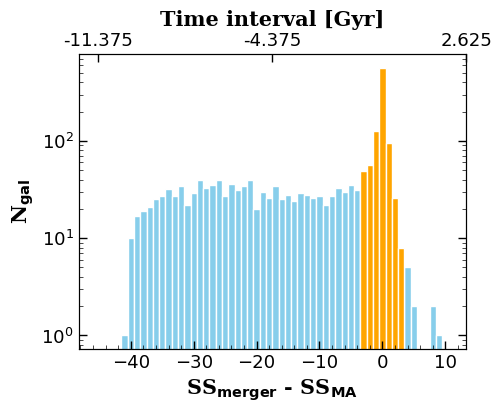}
    \caption{The histograms showing the snapshot interval between the emergence of misalignment ($SS_{\rm MA}$) and the nearest merger event ($SS_{\rm merger}$) in terms of time for all the misaligned galaxies. We use either the beginning or the end point of the merger period whichever is closer to the misalignment. One snapshot interval corresponds to approximately 0.25\,Gyr. The misaligned galaxies near the peak are classified as ``merger-driven'' (orange), and the others are classified as ``non-merger-driven'' (blue). As a result, a total of 928 galaxies are classified as merger-driven misalignment (35\% of the total misaligned galaxies).}
    \label{fig:image_merger_hist}
\end{figure}

Figure~\ref{fig:image_merger_hist} shows the histograms of the time interval between the emergence of misalignment and the merger event for all the misaligned galaxies that experience at least one merger. We use either the beginning or the end point of the merger period whichever is closer to the misalignment. In these histograms, a positive time interval on the X-axis means that the emergence of the misalignment preceded the nearest merger event, which is rare. We have found a peak at $SS_{\rm merger}-SS_{\rm MA}=0$, where the two events occurs at the same time, as we expected. Also, the immediately neighboring bins have significantly more galaxies than the distant bins. 
Note that the size of the bins in the diagram is 0.25\,Gyr, which is the time interval between the gas snapshots in the Horizon-AGN simulation.

We classify the misaligned galaxies around the peak as ``merger-driven'', showing a sign of circumstantial connection between the mergers and misalignment. The misaligned galaxies with snapshot differences equal to or smaller than three (a total of 7 bins) are classified as the merger-driven (orange area) because we do not see a signal of difference in the histogram beyond this range. The duration of time over the 7 snapshots corresponds to 1.75\,Gyr which is comparable to the merging timescale shown in Figure~\ref{fig:image_merger_time}.
%, based on the mean free-fall time. 
Through this classification method, 928 of the total 2,662 (35\%) misaligned galaxies have been identified as merger driven. A different choice of the interval cut, for example 9 or 11 instead of 7 bins, would not lead to any notable difference in the result.

It should also be noted that the right wing from the peak in Figure~\ref{fig:image_merger_hist} is shorter than the left wing. This makes sense in the sense that misalignment likely trails rather than leads a merging event in most cases (see, Figure~\ref{fig:image_seq_merger} for example). Another possible explanation for the uneven wing is the epoch of the sampling. Our galaxies are sampled at $z=0.018$, and so it is impossible to follow their misalignment beyond $z=0$. The presence of some galaxies beyond the peak (the right wing) despite that is because some of the misaligned galaxies formed their misalignment somewhat earlier than $z=0.018$. In this sense, the right wing may not fully represent the true shape of the distribution in this diagram. We may attempt to recover some of the true distribution by considering only the left wing (the three orange bars on the left of the peak) and doubling it. This leads to an upper limit of 39\%, which is somewhat higher than our previous estimate (35\%).

The galaxies with extremely slow rotation (e.g., $V/{\sigma}< 0.2$) must be treated with care. These galaxies are dispersion-dominated systems, and their stellar rotational axes are not well defined. When we exclude the misaligned galaxies with $V/{\sigma}< 0.2$, 711 (30\%) out of the total 2,333 misaligned galaxies are classified as merger driven.

\begin{figure}
	% To include a figure from a file named example.*
	% Allowable file formats are eps or ps if compiling using latex
	% or pdf, png, jpg if compiling using pdflatex
	\includegraphics[width=\columnwidth]{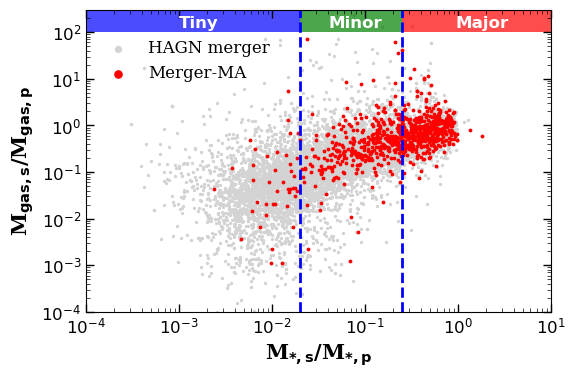}
    \caption{The stellar and the cold gas mass ratio of mergers in Horizon-AGN. All the mergers since $z=0.2$ are marked with gray dots, and the merger-driven misaligned galaxies based on Figure~\ref{fig:image_merger_hist} are marked with red dots. Blue dashed lines show demarcations between major, minor, and tiny mergers.}
    \label{fig:image_merger_mass}
\end{figure}

\subsubsection{Merging mass ratio and misalignment}

When a merger occurs, the merging mass ratio naturally has a significant impact on the rotation of the primary galaxy \citep[e.g.,][]{2018ApJ...856..114C, 2018MNRAS.473.4956L}. As the misalignment is also caused by the rotational axis changes, the mass ratio is thought to have a significant impact on the number of merger-driven misaligned galaxies.

Figure~\ref{fig:image_merger_mass} shows the stellar (X-axis) and the cold gas (Y-axis) mass ratio of mergers in Horizon-AGN. All the mergers since $z=0.2$ are marked with gray dots, and the merger-driven misaligned galaxies seen above are marked with red dots. Among the 928 merger-driven misaligned galaxies, 426 are due to the major merger (46\%), 392 are due to the minor merger (42\%), and 110 are due to the tiny merger (12\%).

The contributions from the major and minor mergers are comparable.
Therefore, considering that minor mergers are more frequent than major mergers in Horizon-AGN by a ratio of 5:2 (see, Section~\ref{sec:hagn_merger}), major mergers can be said to be more effective than minor mergers in generating misalignments by a factor of 5/2, which is understandable.

Minor mergers with a low mass ratio (i.e., tiny mergers) are unlikely to cause notable misalignments.
They may either experience several small mergers at the same time or work together with other (non-merger) mechanisms to create misalignments. Meanwhile, half of the tiny-merger-driven misaligned galaxies (49) have $V/{\sigma} <0.2$ at the time of misalignment formation, whereas only 15\% of the major-merger-driven misaligned galaxies show such low values of $V/{\sigma}$. 
In addition, their median cold gas fraction is only one-third of that of all Horizon-AGN galaxies at $z=0$. Both of these make it very difficult to measure the misalignment of such insignificant mergers reliably.

\subsection{Non-merger-driven misalignment}
\label{sec:origin_non_merger}

In Section~\ref{sec:origin_merger}, we found that about 35\% of the misaligned galaxies originated from mergers. This means that 65\% of the misalignment has been created by ``non-merger'' origins, which is the topic of this section. We investigate non-merger formation channels and classified the non-merger-driven misaligned galaxies into three categories: interaction-driven, group-driven, and secularly-driven misaligned galaxies.

\begin{figure}
	% To include a figure from a file named example.*
	% Allowable file formats are eps or ps if compiling using latex
	% or pdf, png, jpg if compiling using pdflatex
	\includegraphics[width=\columnwidth]{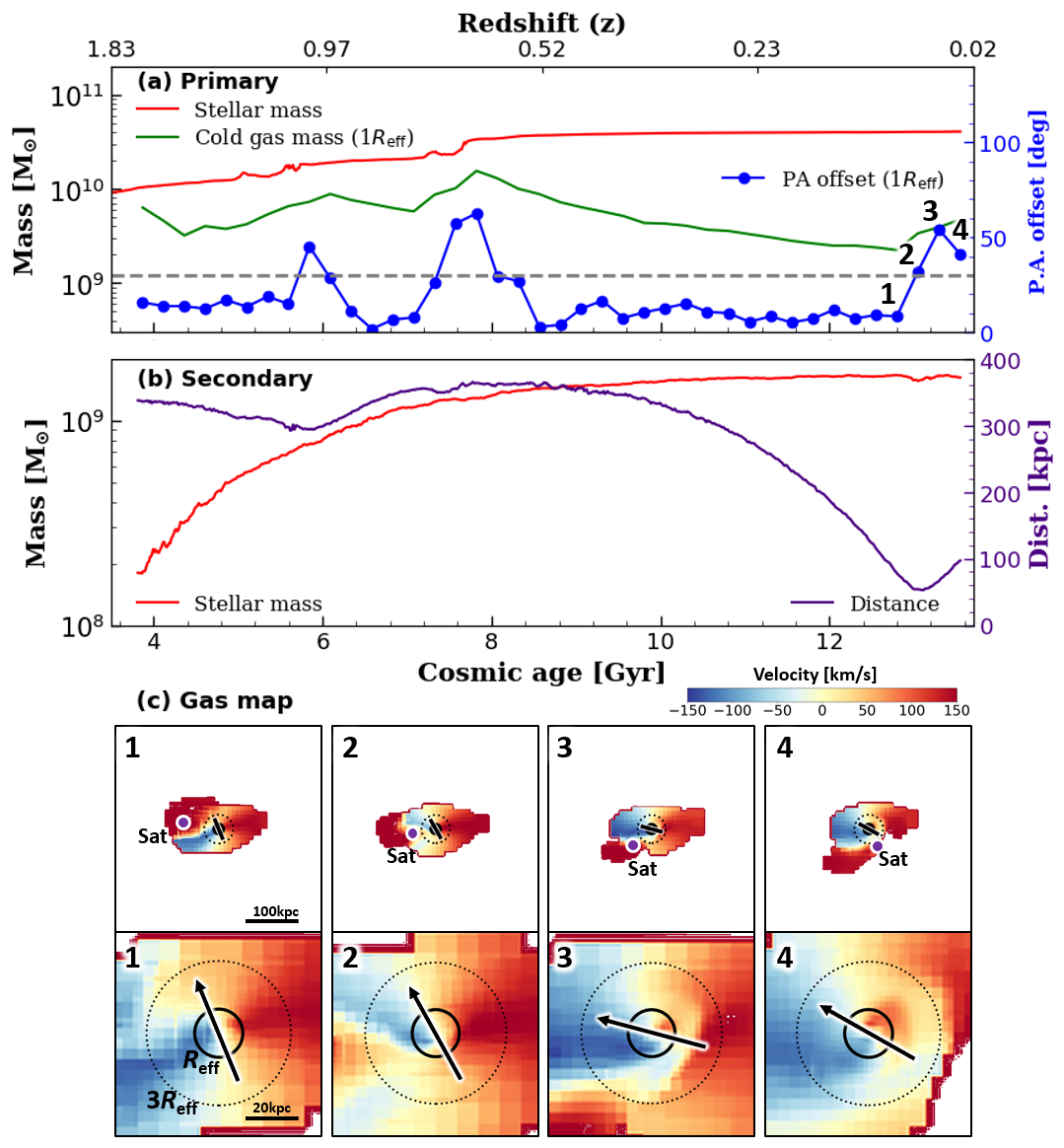}
    \caption{An example of an interaction-driven misalignment formation in a similar format to Figure~\ref{fig:image_seq_merger}. Panel (a): the stellar mass (red), cold gas mass (green), and PA offset (blue) of the primary galaxy. The horizontal dashed line shows $30\,^{\circ}$, which is our criterion for the misalignment. Panel (b): the stellar mass (red) of the secondary galaxy and the distance (violet) from the primary galaxy. Panel (c): cold gas Doppler maps corresponding to each stage and the gas rotational axis at $1\,R_{\rm eff}$. We present two gas maps with different fields of view. The effective radius and $3\,R_{\rm eff}$ are expressed as solid and dotted lines, respectively.}
    
    \label{fig:image_seq_interaction}
\end{figure}

\begin{figure*}
	% To include a figure from a file named example.*
	% Allowable file formats are eps or ps if compiling using latex
	% or pdf, png, jpg if compiling using pdflatex
	\includegraphics[width=\linewidth]{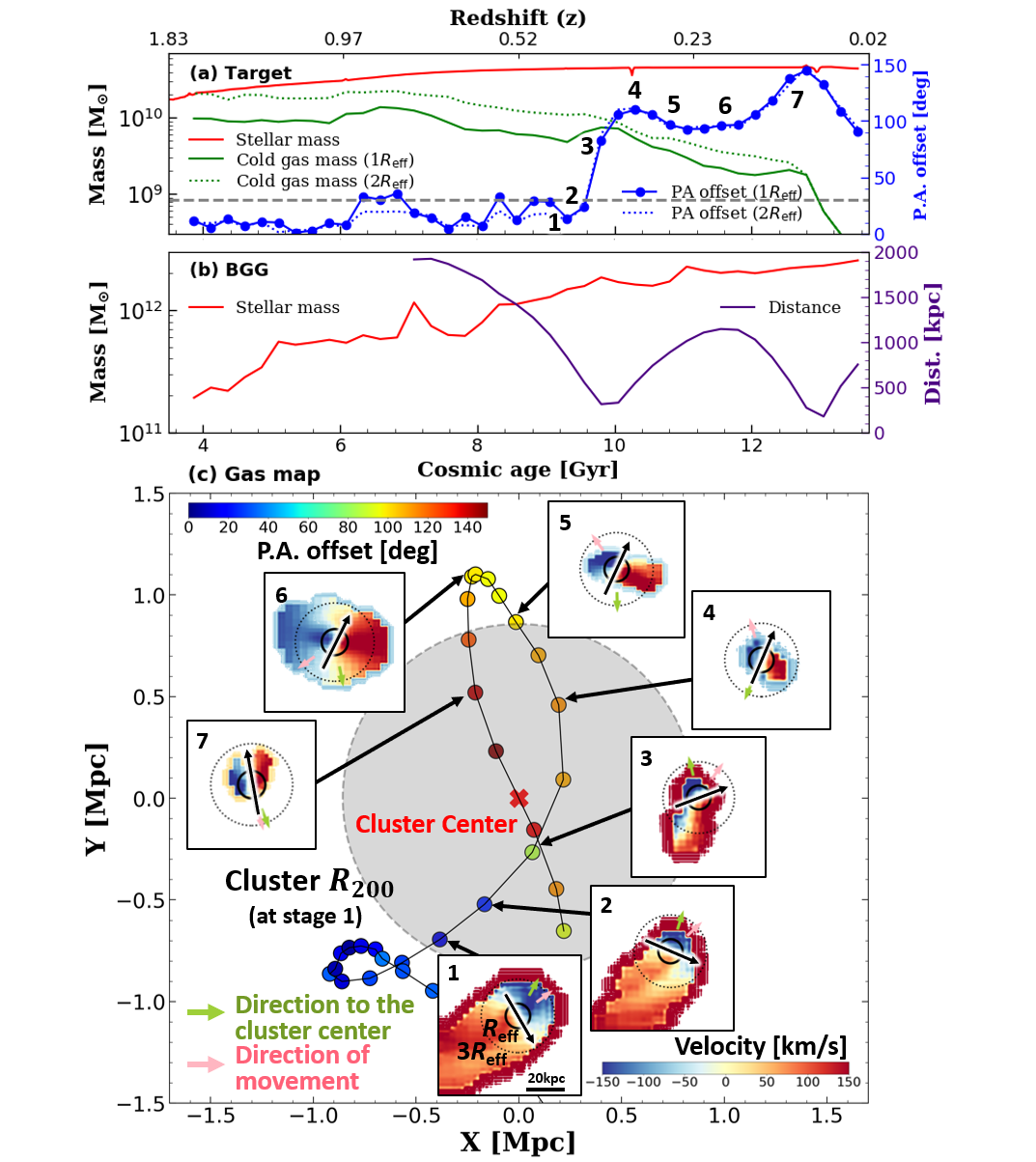}
    \caption{An example of an group-driven misalignment formation in a similar format to Figure~\ref{fig:image_seq_merger}. Panel (a): the stellar mass (red), cold gas mass (green), and PA offset (blue) of the target galaxy. The horizontal dashed line shows $30\,^{\circ}$, which is our criterion for the misalignment. Panel (b): the stellar mass (red) of the BGG and the distance (violet) from the target galaxy. Panel (c): cold gas Doppler maps corresponding to each stage and the infalling trajectory of the galaxy. The virial radius ($R_{\rm 200}$) at Stage 1 is expressed by the gray shade region. Each point represents the position of the galaxy, and is color-coded by the PA offset. The effective radius (solid circle) and $3\,R_{\rm eff}$ (dotted circle) of the galaxy are shown in the inset panels. The green and pink arrows indicate the direction to the cluster center and the galaxy's motion, respectively.}
    \label{fig:image_seq_group}
\end{figure*}

\subsubsection{Interaction-driven misalignment}
\label{sec:origin_inter}
Some galaxies develop the misalignment through interactions with nearby galaxies. The misalignment forms when the gas mass exchange occurs between galaxies mainly through gas accretion, or when the gas velocity field is altered as the gas is affected by nearby galaxies.

Figure~\ref{fig:image_seq_interaction} shows an example of the ``interaction-driven'' misaligned galaxies in a similar format to Figure~\ref{fig:image_seq_merger}. There is the primary (target) galaxy in the center having an aligned gas disk, and the secondary galaxy with a mass of about one-twentieth of the primary galaxy is approaching (Stage 1). The secondary galaxy nears within 50\,kpc and donates its gas to the primary galaxy. The gas from the secondary galaxy, equivalent to the amount of pre-existing gas, flows inside $1\,R_{\rm eff}$ of the primary galaxy. Due to the gas inflow, the gas rotational axis is altered and a misalignment is developed (Stage 2). The misalignment of the primary galaxy continues to grow. However, unlike the merger-driven case, the misalignment is not strong. In particular, the gas in the outer part of the galaxy ($\simeq 3\,R_{\rm eff}$) more or less maintains its original rotation (Stage 3). As a result, the PA offset of the target galaxy inside $1\,R_{\rm eff}$ starts to settle down, i.e., realign (Stage 4). Note that there is little stellar accretion to the target galaxy. Thus, similarly to the minor-merger case, the stellar mass and stellar rotational axis are not affected significantly. 
As discussed in Section~\ref{Mergers and misalignment}, tidal effects besides mass exchanges may also contribute to the formation of the interaction-driven misalignment.

\subsubsection{Group-driven misalignment }
\label{sec:origin_group}
The gas of group member galaxies can be influenced by interactions with the medium of the dense environment \citep[e.g.,][]{1972ApJ...176....1G, 2000Sci...288.1617Q}.
The member galaxies are also affected by the brightest group galaxy (BGG) which is the most influential object in a group.
Therefore, a large number of galaxies develop the misalignment when they pass near the BGG, as we noted in Paper I.

Figure~\ref{fig:image_seq_group} shows an example of ``group-driven'', which is the same galaxy shown in Paper I \citep{Khim+}. As a galaxy infalls to the central region of a group, it loses cold gas by ram pressure stripping and develops the misalignment (Stages 1--4). Most of the group-driven misaligned galaxies develop the misalignment when they pass the pericenter (Stage 4 in the case of the example galaxy). However, some galaxies have a time difference between the misalignment formation and either/both of the epoch of pericenter passes or/and the epoch of the nearest PI maximum. In addition, the PA offsets of group-driven cases are characterized by developing over a longer period of time compared to other cases (e.g., interaction and merger driven). This is understandable because the group-driven process works in the timescale of the orbital (crossing) time, which is much longer inside a group (or cluster) than between two galaxies.

One of the important features of the group-driven is that their gas rotational axis changes significantly, while their stellar rotational axis is virtually fixed. Thus, the group-driven misalignment is mainly due to the changes in the gas rotational axis. The ram pressure stripping can be a possible origin, as mentioned in Paper I \citep{Khim+}.

We have found that the gas mass of some galaxies does not decrease within $1\,R_{\rm eff}$ when they develop the misalignment. Some galaxies even show a temporary increase in their gas mass (a solid green curve in Panel (a)) accompanied by a significant increase in PA offset (Stages 1--3 and 7). This may be puzzling because we can easily expect the strongest candidate for group processes, ram pressure likely reduces the gas fraction, which is not visible in this diagram clearly. 
However, during the same period, their total amount of gas (e.g., within $2\,R_{\rm eff}$, shown as a dotted green curve) decreases steadily. 
This suggests that the gas pushed into $1\,R_{\rm eff}$ by ram pressure or interaction with the BGG may have created the misalignment.

\subsubsection{Secularly-driven misalignment}
\label{sec:origin_secular}

Some galaxies seem to develop their misalignment without any interaction with other objects.
It was difficult to pin down the main physical process that is behind the ``secularly-driven'' case. We find some galaxies showing abrupt motion of gas perhaps due to the episodic gas outflow from star formation or AGN activities. We also find some galaxies whose gas mass increases without a clear reason but apparently from outside probably from neighboring filaments. 
Though very rare, we also find some galaxies accreting gas from the vicinity which is in fact the remnant gas from the previous merger event. It is not feasible to find definite proofs for any of these events mainly because the resolution in time, mass, and space is limited. The Horizon-AGN simulation may not have sufficiently high resolutions for investigating secular processes in detail. In this sense, {\em our ``secularly-driven'' class is essentially ``the rest'' from the other classification categories}.

\subsubsection{Non-merger-driven misalignment classification}

In this section, we classify non-merger-driven misaligned galaxies into the three categories identified above. We explore the short-range effects due to neighboring galaxies and the long-range effects due to the galaxy group itself. After that, we classify the galaxies as to which formation process plays a leading role in developing the misalignment.

We introduce the perturbation index (PI) to investigate the interactions with the nearby galaxies \citep{1990ApJ...350...89B}:
\begin{equation}
    PI=\log \left[ \int_{t_{\rm 0}}^{t} \sum_{i=0}^n \left(\frac{M_{\rm p,i}}{M_{\rm gal}}\right) \times \left(\frac{R_{\rm gal}}{d_{\rm p,i}}\right)^3\, dt/Gyr \right].
	\label{eq:PI}
\end{equation}
In this equation, $M_{\rm gal}$ and $R_{\rm gal}$ represent the mass and the size of the target galaxy, respectively, and $M_{\rm p,i}$ and $R_{\rm p,i}$ are the mass and distance of the i-th perturber, respectively. As we are interested in the rotational axis changes within $1\,R_{\rm eff}$, we have decided to use $R_{\rm eff}$ for the size of a galaxy in the PI measurement.
To calculate PI, all the identified galaxies within 2\,Mpc around the target galaxy are considered as perturbers. Due to the steep (cubic) dependence on distance, the use of a greater limit than 2\,Mpc would not have a large impact on the PI measurement. 
In order to obtain more accurate PI values for 61 coarse gas snapshots, we have measured the PIs for the whole of 787 stellar snapshots and derived the mean values for the 61 gas snapshots, which was used in our analysis.

\begin{figure}
	% To include a figure from a file named example.*
	% Allowable file formats are eps or ps if compiling using latex
	% or pdf, png, jpg if compiling using pdflatex
	\includegraphics[width=\columnwidth]{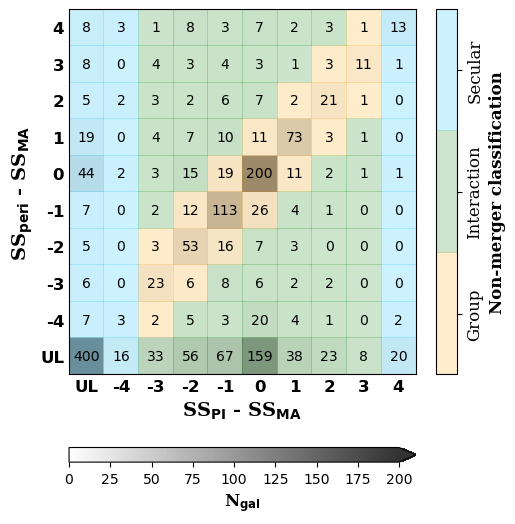}
    \caption{Non-merger-driven misaligned galaxies classification. We draw a two-dimensional histogram by placing $SS_{\rm PI} - SS_{\rm MA}$ on the abscissa and $SS_{\rm peri} - SS_{\rm MA}$, where $SS_{\rm MA}$ is the snapshot number for misalignment formation, $SS_{\rm PI}$ is the snapshot number for the local maximum value of PI, and $SS_{\rm peri}$ is snapshot number for the epoch of the nearest pericenter pass to the BGG. The histogram ranges from –4 to 4, and the galaxies are placed at ``unlinked'' (UL) when the time difference of PI or pericenter peaks cannot be expressed in that range. Each pixel is represented darker as the number of galaxies (the number in each pixel) it contains increased. The background colors present a classification scheme for non-merger-driven misaligned galaxies. We classified non-merger misaligned galaxies into three categories, depending on where they were located.}
    \label{fig:image_nonmerger_hist}
\end{figure}

To investigate the effect of nearby galaxies to the misalignment formation, we perform a similar task to that of Section~\ref{sec:origin_merger_frac}. First, we identify the PI local maxima of each galaxy. To select statistically-significant local maxima only, we consider only the peak values higher than $-5.3$, the median value of PI of the Horizon-AGN galaxies in the local Universe. A different choice of this cut makes only a small difference in the result. For example, the use of the 25\% quantile instead of the median would decrease the fraction of secularly-driven misaligned galaxies by 7.7\%. (572 to 528.) 
Next, we measure the snapshot number for the misalignment formation ($SS_{\rm MA}$) where the PA offset becomes larger than 30\,degrees for the first time. We also measure that of the epoch of the maximum value of PI ($SS_{\rm PI}$). Finally, we compute the difference between the two, which is, $X = SS_{\rm PI} - SS_{\rm MA}$.

We investigate the effect of galaxy groups (and the BGG) in a similar way. The BGG is the most influential object in a group, and its surrounding area would also provide a high degree of ram pressure \citep{2018ApJ...865..156J}. Hence, we use the distance to the BGG as a proxy of the magnitude of the group environmental effect. We measure the difference between the snapshot for the misalignment formation epoch ($SS_{\rm MA}$) and the snapshot for the epoch of the closest {\em pericenter pass} to the BGG in terms of time ($SS_{\rm peri}$), that is, $Y = SS_{\rm peri} - SS_{\rm MA}$.

We then draw a two-dimensional histogram by placing $X$ on the abscissa and $Y$ on the ordinate (Figure~\ref{fig:image_nonmerger_hist}). The range of histograms is set to $\pm4$ snapshots from $SS_{\rm MA}$. If the time interval deviates from the given range ($-4$ to 4), the galaxy is categorized to ``unlinked'' (UL), that is, unlinked to the environmental effects. When either of $X$ or $Y$ is out of the range, the galaxy is added to the grids in the bottom row or left-most column. On the other hand, if both $X$ and $Y$ are out of the range, the galaxy is put to the bottom left grid (coordinate UL-UL). If a galaxy is located in the ``UL'' grids in the left-most column, it means there is little or no causal connection between its misalignment and the environmental effects from nearby galaxies (PI). Similarly, galaxies in the grids in the bottom row show misalignment from something other than the environmental effects caused by the BGG. Note that the galaxies in the field environment, whether merging or not, do not have a value of $SS_{\rm peri}$ and hence would be put into the UL grids on the bottom.

\begin{deluxetable*}{lcccccc}
\tablenum{1}
\tablecaption{The formation channels of the misalignment and their level of contributions at $z=0.018$ \label{tab:messier}}
\tablewidth{100pt}
\tablehead{
\colhead{Origin} 
&\multicolumn{2}{c}{Raw$^{a}$} 
&\multicolumn{2}{c}{Symmetry assumed$^{b}$}
&\multicolumn{2}{c}{$V/{\sigma} \geq 0.2$ $^{c}$}\\
\colhead{} & \colhead{Number} &  \colhead{Fraction$^{d}$} & \colhead{Number} &    \colhead{Fraction} & \colhead{Number} &  \colhead{Fraction} 
}

%\decimalcolnumbers
\startdata
Total & 2662 & 100\% & 2662 & 100\% & 2333 & 100\% \\
\hline
Merger & 928 & 35\% & 1030 & 39\% & 711 & 30\% \\
- Major & 426 & 16\% & 459 & 17\% & 362 & 16\% \\
- Minor & 392 & 15\% & 425 & 16\% & 288 & 12\% \\
- Tiny & 110 & 4\% & 146 & 5\% & 61 & 3\% \\
\hline
Non-Merger & 1734 & 65\% & 1632 & 61\% & 1622 & 70\% \\
- Group & 610 & 23\% &  610 & 23\% & 590 & 25\% \\
- Interaction & 552 & 21\% & 450 & 17\% & 507 & 22\% \\
- Secular & 572 & 21\% &  572 & 21\% & 525 & 23\% \\
\enddata
\begin{tablenotes}
\small
\item{$^{a}$ The raw classification of misalignment formation channels in Horizon-AGN.}
\item{$^{b}$ The result from doubling the left wing of Figure~\ref{fig:image_merger_hist} (see, Section~\ref{sec:origin_merger_frac})}.
\item{$^{c}$ The result from excluding the galaxies with $V/{\sigma}< 0.2$}.
\item{$^{d}$ The fraction shows the number fraction of the galaxies of the total misaligned galaxies. While the binomial standard deviation of each sample is less than 1\%, the fractions are subject to changes depending on the classification details and are uncertain by roughly 10\% in each category.}
\end{tablenotes}
\end{deluxetable*}
\label{table:1}

The darkness of each pixel increases with the number of galaxies in each pixel. A couple of features are outstanding, and these are useful for us to explore the origins of misalignment, as discussed below.

Now, we classify the non-merger-driven misaligned galaxies, based on Figure~\ref{fig:image_nonmerger_hist}. The background colors present a classification scheme for non-merger-driven misaligned galaxies. We first classify secularly-driven misaligned galaxies (the blue area) when $|X|>3$, which developed the misalignment without significant contribution from external (environmental) perturbers. This is similar to what we have described for the classification of merger-driven cases in Section~\ref{sec:origin_merger}. As a result, a total of 572 galaxies are classified as secularly driven (21\% of the total misaligned galaxies).

The galaxies whose misalignment seems linked with the nearest PI maximum, i.e., $|X| \leq 3$, can be divided into two categories depending on whether the galaxy was affected by the BGG. One of the notable feature in the figure is the diagonal distribution of galaxies with their similar values between $X$ and $Y$, which implies that the BGG {\em is} the main driver of both $X$ and $Y$.

The galaxies along this line are close to their BGG and so have a high value of PI, and their misalignment is likely generated by the interaction with the BGG and the central group environment; hence we classify them as group-driven (the orange area).
It should be noted that we widen the diagonal band by one snapshot in both directions in the figure, to consider the poor time resolution of the gas snapshots. We have identified 578 group-driven misaligned galaxies (23\%) using our classification method.

On the other hand, the galaxies away from the diagonal line have their origin for the misalignment from something other than the BGG or the central group environment. 
These galaxies instead develop their misalignment from the strong interaction with nearby galaxies and are hence classified as interaction-driven (the green area). Of the total misaligned galaxies, about 21\% or 584 galaxies, are classified as interaction-driven based on the classification scheme of Figure~\ref{fig:image_nonmerger_hist}.

Some of the interaction-driven misaligned galaxies may have been sampled just before a merging process. If we followed them through beyond $z=0$, they would be classified as merger-driven instead, as was discussed in Section~\ref{sec:origin_merger_frac}. In this sense, the current estimate for the fraction of interaction-driven misaligned galaxies measured at $z=0.018$ is an upper limit. 
The correction to the estimate for the merger-driven misaligned galaxies discussed in Section~\ref{sec:origin_merger_frac} likely affects the estimate for the interaction-driven misaligned galaxies with similar magnitudes but in the opposite direction. In other words, the correction from 35\% to 39\% for the merger driven would require a correction to the interaction driven by a similar magnitude (i.e., from 22\% to 18\%).

As we performed in Section~\ref{sec:origin_merger_frac}, we measure the level of contribution from non-merger-driven misalignment formation channels only for the galaxies with $V/{\sigma} \geq 0.2$. However, excluding the non-merger-driven misaligned galaxies with $V/{\sigma}< 0.2$ does not affect our result significantly.

Table~\ref{table:1} shows a summary of our classification scheme, including merger-driven misaligned galaxies. However, we would like to note that a galaxy may have more than one channel to develop misalignment: for example, interacting galaxies inside a large group or cluster. Nevertheless, we will attempt to estimate the degrees of contribution from various channels/categories in misalignment formation. The estimates of fractions are subject to changes depending on the classification details and are uncertain by roughly 10\% in each category.

\section{Properties of misaligned galaxies}

\label{sec:property}

Misalignment formation processes that change the rotational axes of a galaxy likely affect other galaxy properties as well. Conversely, misalignment processes are critically determined by the environmental status of a galaxy.
In this section, we further investigate the properties of misaligned galaxies.

\begin{figure}
	% To include a figure from a file named example.*
	% Allowable file formats are eps or ps if compiling using latex
	% or pdf, png, jpg if compiling using pdflatex
	\includegraphics[width=\columnwidth]{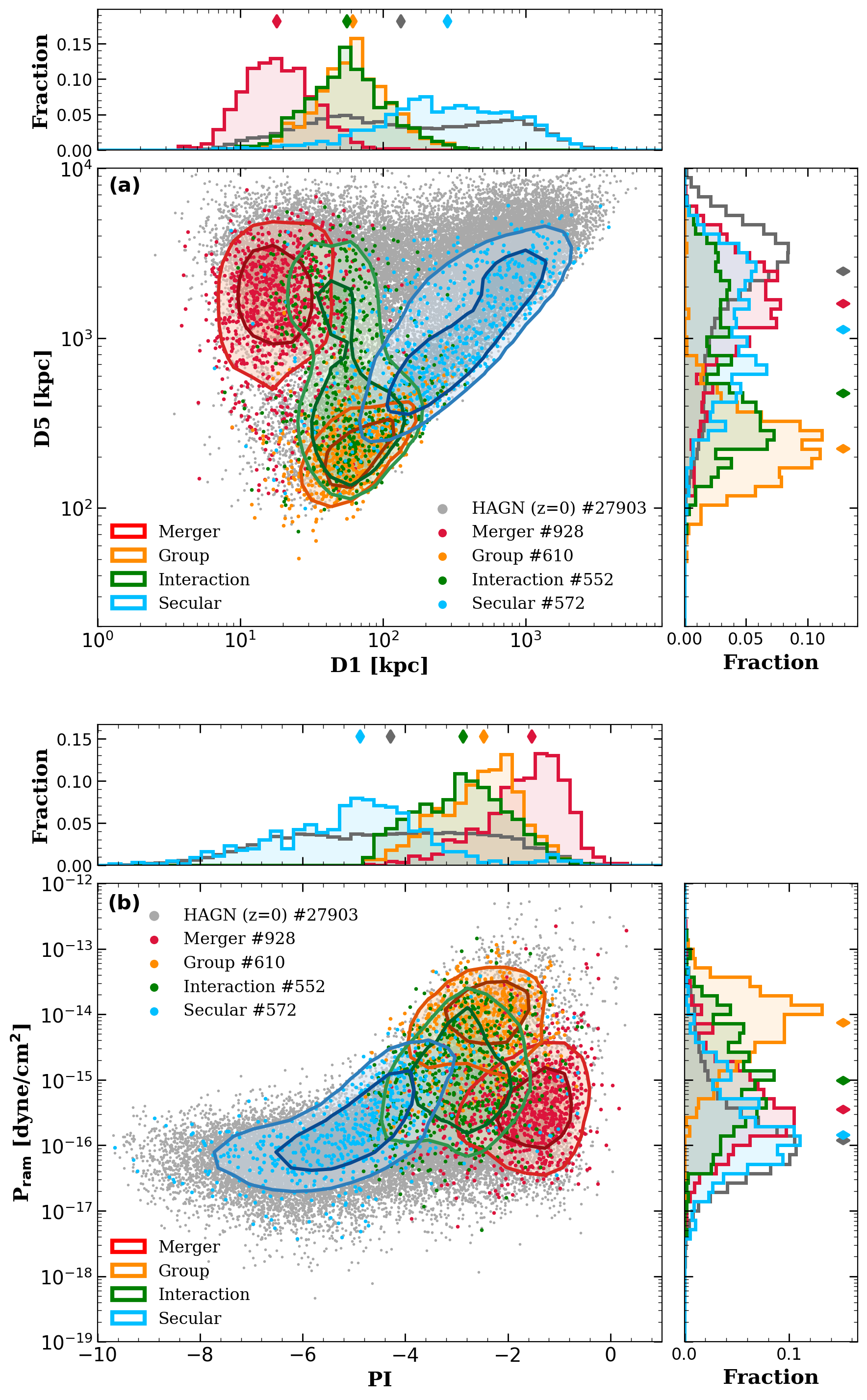}
    \caption{The environmental properties of the misaligned galaxies depending on their origin. Main plot: misaligned galaxies are expressed in different colors corresponding to their origin. The Horizon-AGN galaxies at $z=0.018$ are shown in gray dots. The 0.5 and 1 $\sigma$ contours are also presented. Histograms: histograms for each parameter are shown on the top and right of the figure. Diamond symbols show the median values. Panel (a): D1 against D5, where D1 and D5 are the distance to the closest and the fifth closest galaxy. Panel (b): the perturbation index (PI) against the ram pressure.}
    \label{fig:image_property_env}
\end{figure}

\begin{figure}
	% To include a figure from a file named example.*
	% Allowable file formats are eps or ps if compiling using latex
	% or pdf, png, jpg if compiling using pdflatex
	\includegraphics[width=\columnwidth]{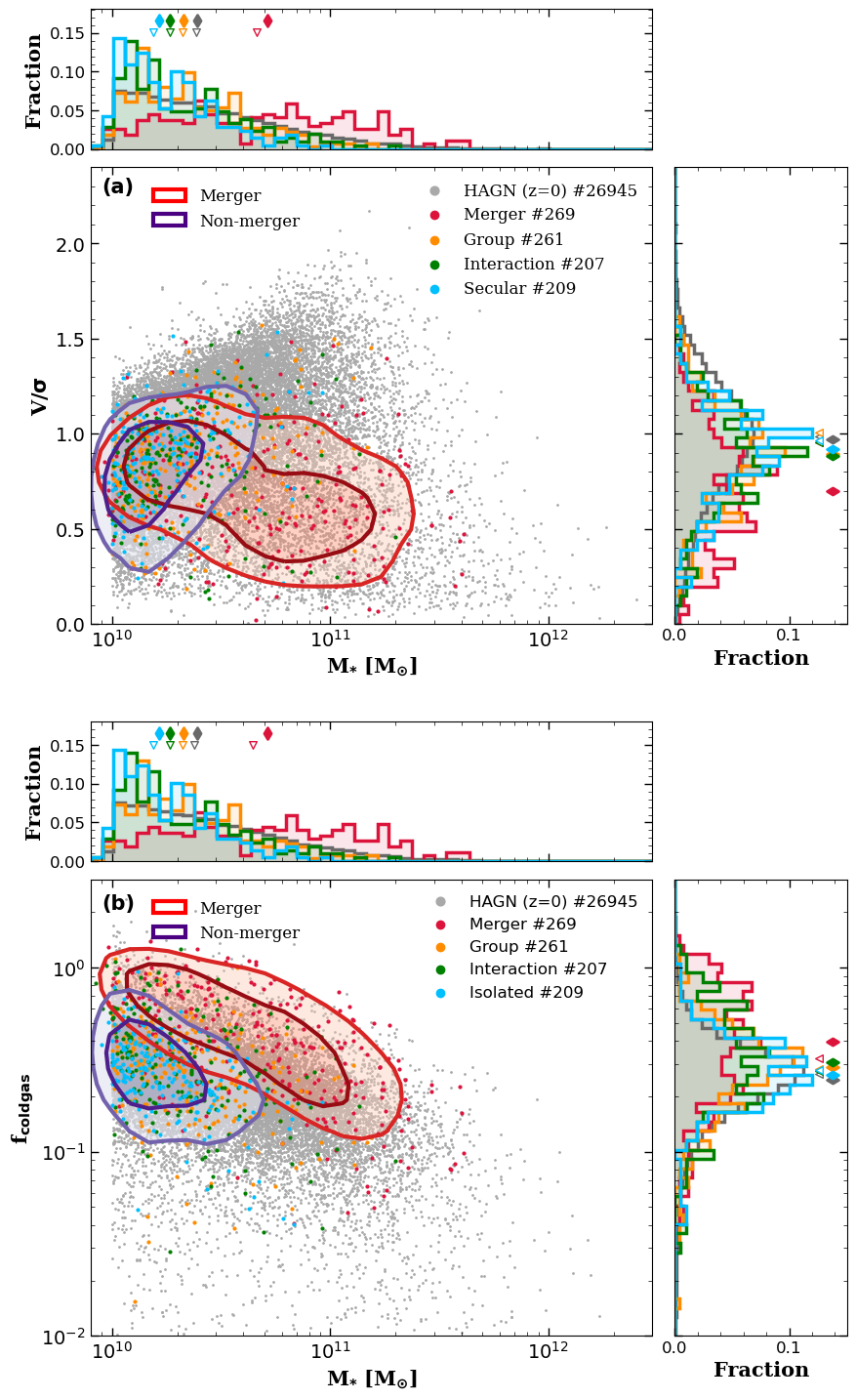}
    \caption{The physical properties of the misaligned galaxies depending on their origin in a similar format to Figure~\ref{fig:image_property_env}. We use only the galaxies that have been aligned for at least 2.5\,Gyr before the misalignment formation. Main plot: the galaxies with the 0.5 and 1 $\sigma$ contours for merger-driven and non-merger-driven are presented. Histograms: diamond symbols show the median values, and triangle symbols show the median value of progenitors (three snapshots ago). Panel (a): $V/{\sigma}$ against stellar mass. Panel (b): the cold gas fraction ($M_{\rm cold\,gas}/M_{\rm *}$) against stellar mass.}
    \label{fig:image_property_phy}
\end{figure}

\subsection{Environmental properties}
\label{sec:property_env}

Figure~\ref{fig:image_property_env} shows the environmental properties of the misaligned galaxies. The misaligned galaxies are classified based on the scheme in Section~\ref{sec:origin} and expressed in different colors corresponding to their origins. We also present their 0.5 and 1 $\sigma$ distributions with color contours.
Meanwhile, all the Horizon-AGN galaxies at $z=0.018$ are shown with gray dots for comparison, regardless of whether they are misaligned. In the side panels, we present the histogram of each parameter. The diamond symbols show the median values.

Figure~\ref{fig:image_property_env}-(a) shows the Horizon-AGN aligned and misaligned galaxies in the plane of D1 and D5, where D1 and D5 are the distance to the closest and the fifth closest galaxy. We use D1 and D5 as a measure of the short-range and long-range environmental density. 
For D5 measurements, we take all the galaxies with mass ratio 1:10 into account while the target galaxies have $M_* > 10^{10}M_{\sun}$.
On the other hand, we use all galaxies without the mass limit for calculating D1 because a very close neighboring galaxy can have a significant impact even if it has a low mass. 
These parameters show a substantial variation from snapshot to snapshot, especially in the case of D1 during merging events. To catch the moment of the maximum environmental effect, we measure D1 and D5 in the target snapshot and the four preceding ones, and take the minimum value.

The merger-driven misaligned galaxies tend to have a lower value of D1 and a higher value of D5 than the others do. A low value of D1 means that there must be another galaxy around for mergers to take place. Similarly, a high value of D5 indicates that mergers are difficult to take place in a very dense environment such as in clusters.
Meanwhile, the secularly-driven misaligned galaxies are rarely affected by either neighboring galaxies or dense environments, so they tend to have high values of D1 and D5.

The group-driven and interaction-driven misaligned galaxies have similar values of D1, lying between the merger-driven and secularly-driven misaligned galaxies. However, their D5 values are somewhat different from each other. The group-driven misaligned galaxies are naturally distributed in dense areas with smaller value of D5, whereas the interaction-driven misaligned galaxies have more diverse values.
Although the interaction-driven misaligned galaxies with a small value of D5 are affected by the group environment, they have been classified as interaction-driven because they developed the misalignment through interactions with non-BGG group member galaxies.

Figure~\ref{fig:image_property_env}-(b) shows the misaligned galaxies in the plane of ram pressure and perturbation index (Section~\ref{sec:origin_non_merger}), which are thought to have a significant impact on the misalignment formation. We measure the degree of ram pressure following the description of \citep{1972ApJ...176....1G}:
\begin{equation}
    P_{\rm ram} = \rho_{\rm \scriptscriptstyle ICM} v_{\rm rel}^2 ,
	\label{eq:ram}
\end{equation}
where $\rho_{\rm ICM}$ is the mass-weighted mean density of ambient gas cells, and $v_{\rm rel}$ is their relative velocity with respect to the galaxy center. We define the ambient gas cells by the surrounding gas cells within 200\,kpc from the galaxy. However, even if we change the aperture criterion, there is no significant impact on the results.

Except for the secularly driven, the misaligned galaxies tend to have both higher values of PI and ram pressure than the aligned galaxies do. Among them, the merger-driven misaligned galaxies have a particularly high value of PI, and the group-driven misaligned galaxies have a particularly high value of ram pressure. The interaction driven have lower values of PI and ram pressure compared to the previous two groups. Finally, the secularly-driven misaligned galaxies tend to have lower values of both PI and ram pressure than other misaligned galaxies do. Their values of PI and ram pressure are closer to those of aligned galaxies than to the other types of misaligned galaxies.

The apparent separation between different groups of galaxies based on their misalignment origins in this diagram is encouraging and supporting our classification schemes presented in Section~\ref{sec:origin}.

\subsection{Physical properties}
\label{sec:property_phy}

In this subsection, we investigate the physical properties of misaligned galaxies depending on their misalignment formation channel. Some galaxies experience more than one misalignment events.
For example, merger-driven misaligned galaxies may have developed another (former) misalignment by interaction with their satellites before the merger in question. Besides, galaxies may have a temporarily-aligned stage during one misalignment event. 
For these misaligned galaxies, it is difficult to investigate how their physical properties are linked with their (most recent) misalignment formation channel. 
Thus, we use in this section only the galaxies that have been aligned for at least 2.5\,Gyr (10 gas snapshots) before the misalignment formation in question.

Figure~\ref{fig:image_property_phy} shows the physical properties of galaxies instead of environmental properties, in a similar format to Figure~\ref{fig:image_property_env}. In the main panel, as all the galaxies with non-merger-driven origins share very similar physical properties, we display their distribution contours together (purple). In the side panels, the histograms for each parameter are separately presented. The diamond symbols show the median values for each group. However, as the physical properties can be changed significantly in the process of developing misalignment, we also measure the properties of their progenitor (three gas snapshots or 0.75\,Gyr ago) and mark their median values with triangles. Thus, we can track the change of the properties of misaligned galaxies by comparing the diamonds and triangles in the histograms.

Figure~\ref{fig:image_property_phy}-(a) shows the stellar mass and kinematic morphology ($V/{\sigma}$) of the galaxies in Horizon-AGN. Note that we have already presented some results from our analysis of the mass and $V/{\sigma}$ ratio of the misaligned galaxies in Paper I. The first to note in this panel is that the merger-driven misaligned galaxies are substantially more massive than the whole sample (gray dots and histograms) and non-merger-driven misaligned galaxies. It is easy to understand because more massive galaxies are more likely to attract neighboring galaxies and experience misalignment.

The mass histogram attached to Figure~\ref{fig:image_property_phy}-(a) also shows that the non-merger-driven misaligned galaxies are all less massive than the whole sample. This is also understandable. The interaction-driven misaligned galaxies are those that are affected by passers-by, and small galaxies are more likely to feel the impact of passers-by more significantly. The group-driven misaligned galaxies are mostly satellites in groups and hence less massive than the whole sample. The isolated galaxies are also known to be less massive than those in dense environments \citep[e.g.,][]{2015ApJS..220....3K}. 

We now focus on the the impact of the misalignment formation on non-merger-driven misaligned galaxies (purple contours). Comparing the diamonds and triangles in the top side histograms, the change in mass in the last three snapshots, that is, during the misalignment process, was negligible. This result is consistent with the previous examples in Section~\ref{sec:origin_non_merger}.
The misalignment process caused $V/{\sigma}$ to decrease slightly due to the new-born ``misaligned'' stars from the misaligned gas disk and the blurring of pre-existing stellar rotation motion ($V/{\sigma}$ histogram).

We then inspect the merger-driven misaligned galaxies. 
They undergo a merger event which resulted in an increase in mass and a decrease in $V/{\sigma}$.
The change in mass is small because the majority of mergers were minor rather than major. The decrease in $V/{\sigma}$ is more pronounced, indicating that the morphology of the galaxy becomes slightly but notably earlier.

Figure~\ref{fig:image_property_phy}-(b) shows the cold gas fraction and stellar mass for all the galaxies. 
There is no significant change in cold gas fraction (between triangles and diamonds) during the process of non-merger misalignment. This may sound strange when it comes to group-driven misaligned galaxies. If ram pressure stripping is the main process that causes misalignment for them, we might naturally expect a decrease in the gas fraction during the misalignment process. As was visible in Figure.~\ref{fig:image_seq_group}-(a), however, the removal of gas due to ram pressure stripping has a time lag from the onset of the misalignment.
As we here use three snapshots {\em before} the snapshot of misalignment for a reference point, the change in the gas fraction appeared negligible in the case of group-driven processes.

On the other hand, the merger-driven misaligned galaxies show a notable difference in the cold gas fraction in terms of the median value and the shape of the distribution.
This trend suggests that for the merger-driven misaligned galaxies, a large amount of the newly-accreted gas overwhelms the pre-existing gas and changes their gas rotational axis, as was demonstrated in Figure~\ref{fig:image_seq_merger}.

Overall, non-merger driven misaligned galaxies do not dramatically change their physical properties during the misalignment formation. However, merger-driven misaligned galaxies increase their stellar and gas masses during the merger event.

\section{Lifetime of the misalignment}
\label{sec:lifetime}

In this section, we investigate how long the misalignment survives. We define the ``lifetime'' of the misalignment as the period from when the misalignment is formed ($\rm{PA\, offset} > 30\, ^{\circ}$) to when the star and gas rotational axes are realigned ($\rm{PA\, offset} \leq 30\, ^{\circ}$). We used the PA offset of 30\,degrees as a criterion for the misalignment in order to compare with the previous papers and observations. The different criteria for the misalignment lead to a different lifetime for misalignment. When we use 15\,degrees instead of 30, for example, the lifetime would become longer, but the overall trend and discussion in this section would be qualitatively the same.

In the previous sections, we have investigated the galaxies at $z=0.018$. However, that sample is not suited for investigating the lifetime of misalignment because the simulation stopped at $z=0$.
Thus, we take a new sample of $z=0.52$ (look-back time of 5.1\,Gyr) to analyze the lifetime of the misalignment. We use the same mass cuts described in Section~\ref{sec:2} ($M_* > 10^{10}M_{\sun}$, $f_g > 3\%$), thus there are 26,411 galaxies of which 3,339 (12.6\%) are misaligned. 
Among the misaligned galaxies, we consider only 2,177 galaxies in the following analysis because these galaxies survive through $z=0.018$ and have a measurable amount of gas.

\begin{figure}
	% To include a figure from a file named example.*
	% Allowable file formats are eps or ps if compiling using latex
	% or pdf, png, jpg if compiling using pdflatex
	\includegraphics[width=\columnwidth]{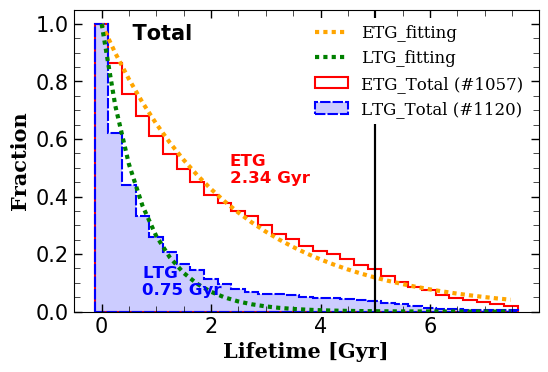}
    \caption{The decay histogram of the misalignment. Based on the misaligned galaxies at $z=0.52$, the histogram shows the number fraction of the galaxies maintaining their misalignment as a function of time. We use an exponential decay function to fit the histogram. This is performed for the range 0--5\,Gyr (the vertical line) of the lifetime only. The orange and green curves show the best-fit of ETGs and LTGs, respectively.}
    \label{fig:image_lifetime_morp}
\end{figure}

\subsection{Properties of galaxies and misalignment lifetime}
\label{sec:lifetime_prop}

Once a galaxy develops the misalignment, the stellar and gas disks attract each other and eventually realign. Some physical properties of the galaxy may influence the speed of the ``settling down'' process: e.g., ellipticity \citep{2019MNRAS.483..458B}.

We first examine the lifetime of the misalignment depending on their kinematic morphology. 
Figure~\ref{fig:image_lifetime_morp} is a decay histogram of star-gas misalignment. The abscissa shows the lifetime of the misalignment, and each bin shows the number fraction of the galaxies maintaining their misalignment as a function of time. 
We find that a considerable number of galaxies maintain their misalignment for more than a few billion years. In particular, a total of 138 galaxies maintain their misalignment through the last gas snapshot ($z=0.018$), and so their lifetime cannot be fully measured.

We divide the galaxies into ETGs and LTGs (see, Section~\ref{sec:hagn_gal}) and investigate their lifetime. Both the histograms of ETGs and LTGs show that the number of misaligned galaxies decays gradually and steadily with time.
However, there is a clear difference between ETGs and LTGs: ETGs tend to have a longer lifetime than LTGs. For example, during the first one gas snapshot interval ($\simeq$ 0.25\,Gyr), about 90\% of the misaligned ETGs still maintain their misalignment, whereas only about 60\% of misaligned LTGs do.

\begin{figure}
	% To include a figure from a file named example.*
	% Allowable file formats are eps or ps if compiling using latex
	% or pdf, png, jpg if compiling using pdflatex
	\includegraphics[width=\columnwidth]{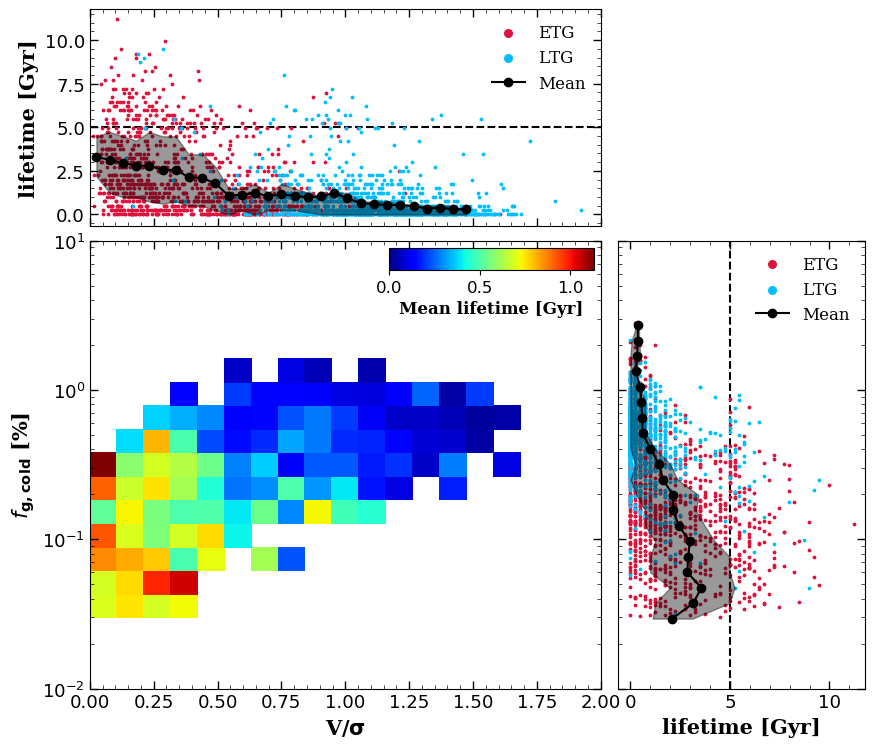}
    \caption{The mean lifetime of the misalignment as a function of the kinematic morphology of the galaxy (X-axis) and the cold gas fraction (Y-axis). Main: each pixel is color-coded by the mean lifetime of the misalignment. Each pixel contains at least five galaxies to ensure statistical significance. Side panels: the lifetime as a function of each parameter. ETGs and LTGs are expressed in red and blue, respectively. The black curve and shaded area show the median lifetime and 25--75 percentiles, respectively. The 5\,Gyr line aforementioned is marked with a dotted line.}
    \label{fig:image_lifetime_gas}
\end{figure} 

To quantify the lifetime of the misalignment, we fit the histogram using an exponential decay function:
\begin{equation}
    N(t) = N_{\rm 0} \exp(-t/\tau),
	\label{eq:decay}
\end{equation}
where $N_{\rm 0}$ is the initial number of misaligned galaxies, $t$ is time, and $\tau$ is the exponential decay timescale. The fit is performed for the range of 0--5\,Gyr (the vertical line) of the lifetime only, because 138 galaxies maintain their misalignment through the end of the simulation as mentioned above.
The decay timescale from the best-fit is 2.34\,$\pm$\,0.05\,Gyr and 0.75\,$\pm$\,0.04\,Gyr for ETGs (orange curve) and LTGs (green curve), respectively. This means that the misalignment lifetime of ETGs is 3.4 times longer than that of LTGs.

Let us extend the analysis further. Paper I demonstrated that misalignment phenomena (in terms of the number fraction and PA offset) are affected by both the kinematic morphology and the cold gas fraction of galaxies. Thus, we investigate whether these parameters have an impact on the lifetime, too. We draw a two-dimensional histogram to analyze the effects of each parameter separately, as shown in Figure~\ref{fig:image_lifetime_gas}. In the main panel, each pixel is color-coded by the mean lifetime of the misalignment. Each pixel contains at least five galaxies to ensure statistical significance. In the side panels, we present the lifetime as a function of each parameter.

\begin{figure}
	% To include a figure from a file named example.*
	% Allowable file formats are eps or ps if compiling using latex
	% or pdf, png, jpg if compiling using pdflatex
	\includegraphics[width=\columnwidth]{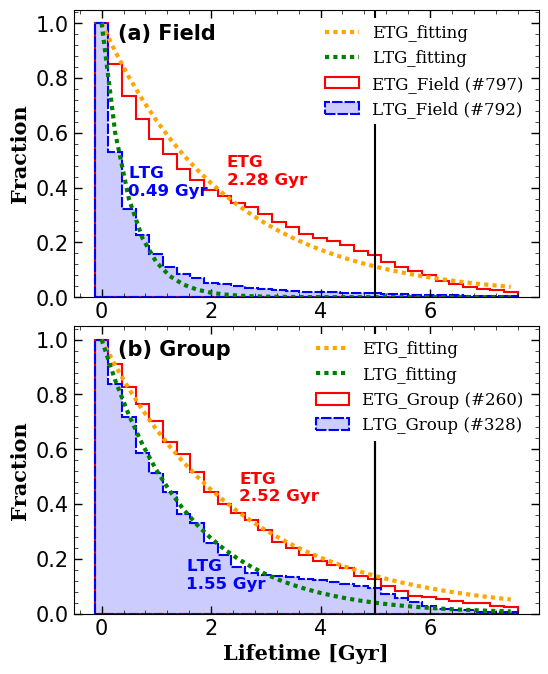}
    \caption{The decay histogram of the misalignment depending on environments in a similar format to Figure~\ref{fig:image_seq_merger}. Panel (a): field galaxies (non-group members). Panel (b): group member galaxies.}
    \label{fig:image_lifetime_env}
\end{figure}

We find in the top panel that the lifetime varies depending on $V/{\sigma}$ even within the same classification of morphology. Overall, the galaxies of earlier-type (i.e., lower values of $V/{\sigma}$) tend to have a longer lifetime of misalignment. This trend is consistent with the result of \cite{2019MNRAS.483..458B}, considering that there is a clear positive trend between the kinematic morphology and ellipticity.

Next, we examine the impact of the cold gas fraction. We find that the relatively gas-poor galaxies tend to have a longer lifetime of the misalignment than gas-rich galaxies. While the gas fraction and $V/\sigma$ (morphology) are closely linked with each other, we find that they independently affect the misalignment fraction. The smaller the two parameters are, the longer the lifetime is. This result is also consistent with the trend shown in the gas-morphology histogram mentioned in Paper I.
We present a simple dynamical model that explains the effect of gas fraction in Appendix \ref{sec:AppA}.

We can explain the impact of the amount of gas in the same way as mentioned in Paper I.
In a galaxy with a higher cold gas fraction, the momentum dissipation due to the interaction between stars and gas is expected to be more efficient. 
This may impact on the settling-down process and decrease PA offset more quickly. Also, star formation in the gas disk may reduce the misalignment lifetime. The new stars born with the kinematic characteristics of the misaligned gas disk are added to the existing stars. As a result, the stellar rotational axis (the direction of the mean angular momentum) would be gradually changed to that of the gas disk.

\subsection{Environment and misalignment lifetime}
\label{sec:lifetime_env}

% Also, the demarcation line between ETG and LTG are marked with a black dashed line.
As noted in Paper I, simulation galaxies show an enhanced misalignment fraction in dense environments, such as massive groups or clusters. Also, we have found that dense environments are closely linked with the misalignment formation (Section~\ref{sec:origin_group}). 
Thus, the galaxies in different environments may have different misalignment lifetimes.

In order to investigate the possible impact of the environment on the lifetime, we divide the galaxies into group members and field (non-group) galaxies (see Section~\ref{sec:hagn_gal}) and measure their lifetime, as shown in Figure~\ref{fig:image_lifetime_env}.
Panel (a) shows the decay histogram in the field environment. The decay timescale is 2.28\,$\pm$\,0.08\,Gyr in ETGs and 0.49\,$\pm$\,0.02\,Gyr in LTGs: 4.7 times different. In contrast, in the dense environment (Panel (b)), the decay timescale is 2.52\,$\pm$\,0.03\,Gyr in ETGs and 1.55\,$\pm$\,0.04\,Gyr in LTGs, leading to the 1.62 times difference. Overall, the decay timescale of misalignment is longer in denser environments. Also, the lifetime of LTGs seems to be affected by the environmental effect more significantly.

Figure~\ref{fig:image_lifetime_env} can be explained by Figure~\ref{fig:image_lifetime_gas}. The group/cluster member galaxies are affected by their environments; member galaxies in groups more easily become early-types \citep[morphology-density relation,][]{1980ApJ...236..351D}, passive in terms of star-formation \citep{2003ApJ...584..210G}, and gas-poor due to the gas stripping process \citep[e.g.,][]{1972ApJ...176....1G, 2000Sci...288.1617Q}.
The infalling ETGs in dense environments steadily lose their cold gas due to the environmental effect. The mean gas fraction when the group ETGs generate the misalignment is 0.27, and it becomes 0.12 after 2.5\,Gyr. The group ETGs move to the bottom of Figure~\ref{fig:image_lifetime_gas} and have a longer lifetime than field ETGs.

The infalling LTGs also lose their cold gas (0.52 to 0.25), as ETGs do. Moreover, they become earlier in kinematic morphology (the mean value of $V/{\sigma}$ of 1.55 to 1.11 after 2.5\,Gyr). They move to the left and bottom in Figure~\ref{fig:image_lifetime_gas} and have a much longer lifetime than field LTGs.
In addition, the misaligned galaxies in group environments often regenerate the misalignment when the galaxy passes the pericenter multiple times (e.g., Stage 7 in Figure~\ref{fig:image_seq_group}). This helps the misalignment of galaxies survive longer than in the field.

\subsection{Origins and misalignment lifetime}
\label{sec:lifetime_ori}

In this section, we explore the lifetime of the misalignment depending on their origins. We classify the Horizon-AGN galaxies at $z=0.52$ with their origins in the same way as in Section~\ref{sec:origin}. Among the 2,177 galaxies, there are 1,313 (60\%) merger-driven, 195 (9.0\%) group-driven, 247 (11\%) interaction-driven, and 422 (19\%) secularly-driven misaligned galaxies. 

\begin{figure}
	% To include a figure from a file named example.*
	% Allowable file formats are eps or ps if compiling using latex
	% or pdf, png, jpg if compiling using pdflatex
	\includegraphics[width=\columnwidth]{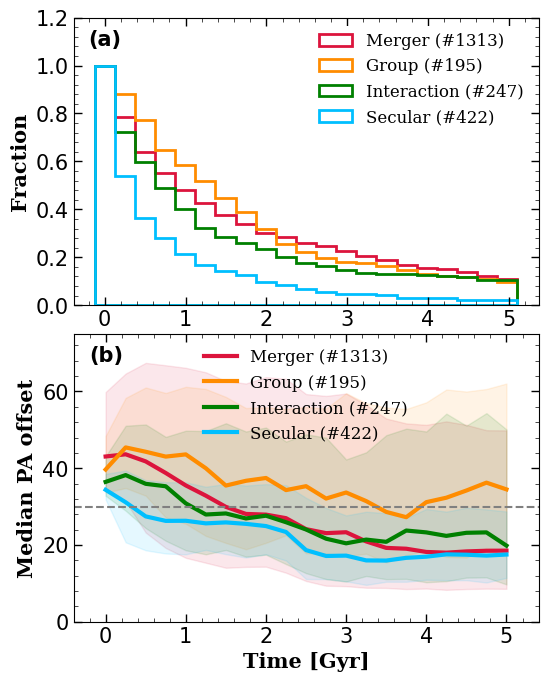}
    \caption{The realigned process of the misaligned galaxies depending on their origins. Panel (a): the decay histogram of the misalignment in a similar format to Figure~\ref{fig:image_lifetime_morp}.
    Panel (b): the median value of the PA offset as a function of time. The colored curves and shaded areas show the median PA offsets and 25--75 percentiles, respectively. The criterion for misalignment ($30\,^{\circ}$) is marked with a horizontal dashed line. The median PA offset for aligned galaxies is $8.8\,^{\circ}$.} 
    \label{fig:image_lifetime_origin}
\end{figure}

There is significant discrepancy between the results here at $z=0.52$ and at $z=0.018$ from Section~\ref{sec:origin} in terms of the number fraction of the merger driven. 
A small part of the discrepancy can be explained by the tendency of higher merger frequencies in the earlier Universe. However, the main reason for the difference is the bias from the sample selection. As we mentioned, we have used only the galaxies that survived through $z=0.018$. A larger fraction of the lower mass galaxies disappeared through galaxy mergers between $z=0.5$ and $z=0.018$. As the merger-driven misaligned galaxies are heavily biased toward massive galaxies, this leads to an increase of the fraction of merger-driven misaligned galaxies, compared to that of $z=0.018$.
However, it should be noted that the difference in the number fractions cannot affect the subsequent analysis because we will measure the lifetime for each origin.

Figure~\ref{fig:image_lifetime_origin}-(a) shows the decay histogram of the misalignment in a similar format to Figure~\ref{fig:image_lifetime_morp}.
Overall, the group-driven misaligned galaxies tend to maintain their misalignment longer than the others, followed by the merger driven and interaction driven. The misalignment decays most rapidly in the secularly driven.

We have found that the non-merger-driven misaligned galaxies share similar physical properties in terms of stellar mass, morphology, and gas fraction when they develop the misalignment (Figure~\ref{fig:image_property_phy}). Thus, the lifetime difference between the three subgroups of the non-merger-driven misalignment is difficult to explain using the information in Figure~\ref{fig:image_lifetime_gas} solely.

Figure~\ref{fig:image_lifetime_origin}-(b) shows the median PA offset as a function of time. The criterion for misalignment (30\,degrees) is marked with a horizontal dashed line. Except for the initial value of PA offset ($x=0$), the group driven tend to have a higher PA offset than other types. Their PA offset decreases gradually due to the environmental effect (Section~\ref{sec:lifetime_env}). 
In the case of the merger driven, on the other hand, their initial PA offsets are similar to those of the group driven but show a rapid decline. The interaction driven, like the merger driven, show a decline in their PA offset. However, since their initial PA offsets are typically less than those of the merger driven, the time for their PA offsets to reach below our PA offset cut for misalignment (30\,degrees) is shorter.
Finally, the secularly-driven misaligned galaxies have the lowest initial PA offsets on average, and so their misalignment dissipates most quickly.

\begin{figure*}
	% To include a figure from a file named example.*
	% Allowable file formats are eps or ps if compiling using latex
	% or pdf, png, jpg if compiling using pdflatex
	\includegraphics[width=\linewidth]{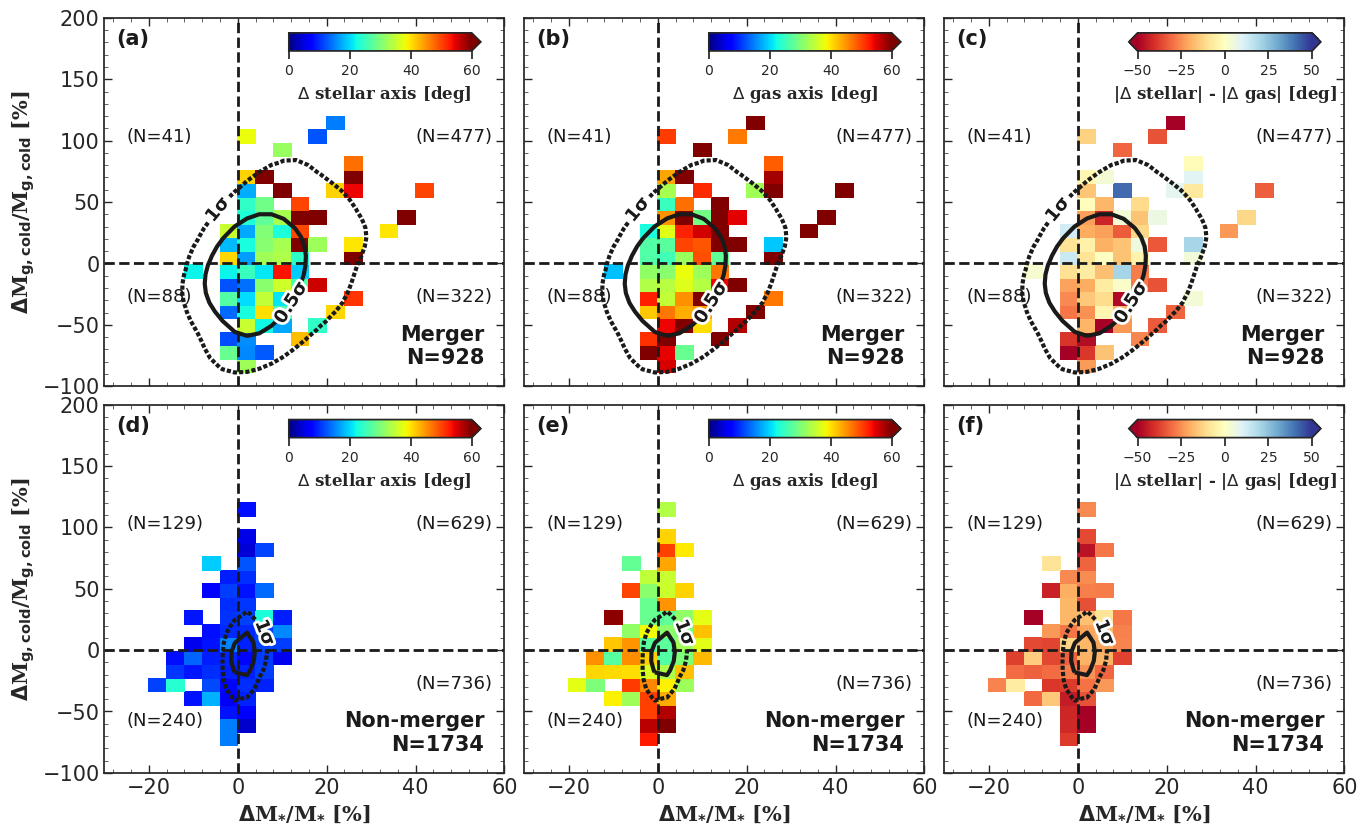}
    \caption{Same as Figure~\ref{fig:image_massmass}, but for the misaligned galaxies at $z=0.018$. Each colored pixel contains at least three galaxies. The number in each quadrant shows the number of galaxies in the region. Note that Panels (c) and (f) have different color scales compared to that of Figure~\ref{fig:image_massmass}.}
    \label{fig:image_massmass_origin}
\end{figure*}

\subsection{Summary on the lifetime}

The results of the lifetime of the misalignment analysis using Horizon-AGN galaxies at $z=0.52$ can be summarized as follows.
First, we have investigated the lifetime depending on the physical properties of galaxies in Section~\ref{sec:lifetime_prop}. 
Morphology ($V/{\sigma}$) and gas fraction are independently anti-correlated with the decay timescale of misalignment (2.28\,Gyr for ETGs and 0.49\,Gyr for LTGs). 
This result is consistent with the misalignment scheme and the discussion of the impact of the gas fraction to the re-alignment process mentioned in Paper I.
Next, we have found that the galaxies in dense environments tend to have a longer lifetime than field galaxies (Section~\ref{sec:lifetime_env}). This tendency seems to reflect the result of Section~\ref{sec:lifetime_prop}. Since infalling galaxies in dense environments lose their cold gas and become earlier in morphology, they tend to have a longer lifetime than field galaxies.
Lastly, we have found that the lifetimes of the misalignment are longer in order of group-driven, merger-driven, interaction-driven, and secularly-driven misalignment in Section~\ref{sec:lifetime_ori}. 
While the different properties of the misaligned galaxies depending on their origin can partly explain this, the lifetime difference among the three subgroups of the non-merger-driven misalignment can be explained by the environmental effect and their initial PA offsets.

%%%%%%%%%%%%%%%%%%%%%%%%%%%  Section 4  %%%%%%%%%%%%%%%%%%%%%%%%
\section{Discussion}
\label{sec:discussion}

\subsection{Main mechanisms of misalignment formation}

\label{sec:dis_env} % used for referring to this section from elsewhere

In this section, we examine and discuss how the misalignment develops. We show the stellar and gas rotational axis changes of the misaligned galaxies at $z=0.018$ by measuring the rotational axes at the snapshot of misalignment formation and one snapshot before. Figure~\ref{fig:image_massmass_origin} shows this result in a similar format to Figure~\ref{fig:image_massmass}. We divide the galaxies into the merger-driven (the upper panels) and non-merger-driven (the lower panels) misaligned galaxies. Each pixel contains at least three galaxies to ensure statistical significance.

The merger-driven misaligned galaxies tend to show a significant increment in their stellar mass due to a major or minor merger. The distribution of red pixels in Panel (a) shows that the greater change in stellar mass is accompanied by the greater change in the stellar axis. The merger driven also show the changes in the cold gas mass due to the merger events. These gas mass changes are also linked with the changes in the gas rotational axis, as shown in Panel (b). Some of the merger-driven misaligned galaxies accrete cold gas from the merging galaxies resulting in the changes in their gas rotational axis. On the other hand, we have also found that some galaxies lose their cold gas due to the AGN or stellar feedback after the merger. Although small in number, some galaxies do not show a change in their gas or stellar mass during one gas snapshot interval (i.e., those near the origin in this panel); yet, their gas rotational axis is still altered by more than 20\,degrees. 
Overall, the misalignment of the merger driven misaligned galaxies is associated with mass changes in both the stars and gas. 
However, Panel (c) shows that most of the pixels are red, which indicates that the rotational axis changes are usually greater in the gas rather than in stars. Thus, the star-gas misalignment of the merger-driven misaligned galaxies is mainly due to the change in the gas rotational axis during mergers.

The non-merger-driven misaligned galaxies show little accretion of stellar mass, unlike the merger-driven misaligned galaxies. Some of them even lose their stellar mass due to tidal stripping. Moreover, they show only the minute changes in their stellar rotational axes when they develop the misalignment, as shown in Panel (d). This means that the examples we have investigated in Section~\ref{sec:origin_non_merger} (e.g., Figures~\ref{fig:image_seq_interaction} and \ref{fig:image_seq_group}) are not exceptional cases.
On the other hand, the gas mass of the non-merger driven misaligned galaxies can be changed due to the interactions with nearby galaxies or environments. The changes in their gas mass lead to the changes in the gas rotational axis, as shown in Panel (e). Like the merger-driven case, the gas rotation axes have been changed at least 20\,degrees even if the galaxies do not show changes in their gas mass. Lastly, Panel (f) indicates that the rotational axis changes of gas outweigh those of stars when the non-merger-driven misaligned galaxies develop the misalignment.

We conclude that the star-gas misalignment of the galaxies is mainly due to the change in the gas rotational axis regardless of its origin.

\subsection{Merger driven fraction}

In Section~\ref{sec:origin_merger_frac}, we have found that about 35\% of the total misaligned galaxies at $z=0.018$ are merger driven. It means that mergers play a crucial role in the star-gas misalignment formation \citep[e.g.,][]{1990ApJ...361..381B, 1991Natur.354..210H, 1996ApJ...471..115B, 1998ApJ...499..635B, 2001Ap&SS.276..909P, 2009MNRAS.393.1255C}.

The misalignment study of \citet{2019ApJ...878..143S} based on the Illustris simulation reported that mergers with merging mass ratios larger than 1:10 did not significantly contribute to the misalignment formation, which may appear contradictory from our result.
The difference comes from the use of different mass ranges for the galaxy sample selection. \cite{2019ApJ...878..143S} used galaxies in the range of $M_* = 2\times 10^9M_{\sun}$ to $ M_* = 5\times 10^{10}M_{\sun}$ in order to investigate the galaxies that are expected to have a low impact of mergers.
On the other hand, we did not set an upper limit on stellar mass, thus our sample included a significant number of massive galaxies. Figure~\ref{fig:image_property_phy} well illustrates that the median value of the stellar mass of the merger-driven misaligned galaxies is roughly $8\times 10^{10} M_{\sun}$, which exceeds the mass range of \cite{2019ApJ...878..143S}.

\begin{figure}
	% To include a figure from a file named example.*
	% Allowable file formats are eps or ps if compiling using latex
	% or pdf, png, jpg if compiling using pdflatex
	\includegraphics[width=\columnwidth]{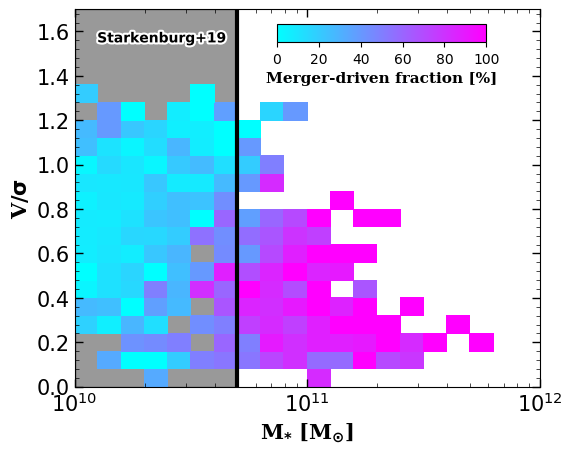}
    \caption{The number fraction of the merger-driven misaligned galaxies. \cite{2019ApJ...878..143S} used galaxies in the range of $M_* = 2\times 10^9M_{\sun}$ to $ M_* = 5\times 10^{10}M_{\sun}$ and the upper limit is marked with a vertical solid line. While the majority of the massive ($M_* \geq 5\times 10^{10}M_{\sun}$) misaligned galaxies are the merger driven, the area shared with their sample selection (a shaded region) is dominated by the non-merger driven.}
    \label{fig:image_merger_frac}
\end{figure}

We plot Figure~\ref{fig:image_merger_frac} to visualize the discrepancy more clearly. In this two-dimensional histogram, each pixel is color-coded by the number fraction of merger-driven misaligned galaxies. The vertical line shows the upper limit of the mass range in \cite{2019ApJ...878..143S}, and the shaded region shows where the two samples overlap.
The majority of the massive misaligned galaxies above the mass range of the sample of \cite{2019ApJ...878..143S} are merger driven (magenta). On the other hand, Horizon-AGN also shows that the low-mass shaded region is dominated by the non-merger-driven misaligned galaxies (cyan), such as interaction-driven, group-driven, and secularly-driven misaligned galaxies. Therefore, the two different cosmological simulations are not conflicting with each other.

%%%%%%%%%%%%%%%%%%%%%%%%%%%  Section 5  %%%%%%%%%%%%%%%%%%%%%%%%
\section{Summary and conclusion} 
\label{sec:5}

The Horizon-AGN simulation contains a large number of galaxies of various properties in various environments and thus offers clear advantages for studying the star-gas misalignment. In this paper, we have explored the origins of the star-gas misalignment in Horizon-AGN and quantified the level of contribution from each formation channel. We have also investigated the properties and lifetime of the misaligned galaxies. Our main results can be summarized as follows.

\begin{itemize}[wide, labelwidth=!, labelindent=0pt]

 \item{We have found that there are four main formation channels of the star-gas misalignment. (i) Merger-driven: the mergers provide a significant amount of stars and gas to the target galaxy and develop a misalignment. (ii) Interaction-driven: the interaction with nearby galaxies also has an impact on the gas rotational axis of the target galaxy. (iii) Group-driven: infalling galaxies in dense environments can be misaligned. Their misalignment is likely formed by the interaction with the BGG or the intracluster medium in the central group environment. (iv) Secularly-driven: some galaxies develop the misalignment without any interaction with other objects. Some of them seem to be misaligned due to the infalling gas from the neighboring filaments.
}

  \item{We have plotted the misaligned galaxies in the plane of D1 and D5, where D1 and D5 %are the distance to the closest and the fifth closest galaxy.
  are a short-range and a long-range environmental density indicator, respectively. The four classifications above are clearly separated out on the D1-D5 plane, hence supporting our classification scheme. They are also clearly separated out in the plane of perturbation index vs. ram pressure.
}

  \item{We have presented the degree of contribution from each misalignment formation channel in terms of number fraction in Horizon-AGN at $z=0.018$ as follows: merger-driven (35\%), group-driven (23\%), interaction-driven (21\%), and secularly-driven (21\%). The estimates of fractions are subject to changes depending on the classification details and are uncertain by roughly 10\% in each category. 
}

  \item{The changes in the rotational axis seem to be linked with the changes in the stellar/gas mass of a galaxy. Since merger-driven misaligned galaxies tend to show significant changes in both stellar and gas masses, both axes can be changed during the merger. However, misalignment is generally more due to the the rotational axis change in gas rather than that in stars. The non-merger-driven misaligned galaxies show little accretion of stellar mass, thus show only the minute changes in their stellar rotational axes. However, their gas mass are changed more significantly due to the interaction with nearby galaxies or environments, which leads to the changes in the gas rotational axis. Thus, the star-gas misalignment of the galaxies is mainly due to the change in the gas rotational axis regardless of its origin.
}

 \item{The decay timescale of the misalignment is strongly linked with the kinematic morphology ($V/{\sigma}$) of the galaxy: early-type galaxies (2.28\,Gyr) tend to have a longer misalignment lifetime than LTGs (0.49\,Gyr). We also found that the morphology and cold gas fraction are both and independently anti-correlated with the misalignment lifetime. }

 \item{
 Galaxies in dense environments tend to have a longer lifetime than field galaxies.
 Also, the lifetime of LTGs seems to be affected by the environmental effect more significantly than that of ETGs: the decaying timescale of ETG and LTG shows 4.7 times and 1.6 times difference in the field and group environments, respectively.
 This tendency seems to reflect the relation between the lifetime and the physical properties of the galaxy. Since infalling galaxies in group environments lose their cold gas and become earlier in morphology, their misalignment tends to have a longer lifetime than that of field galaxies.
 We also found that the lifetimes of the misalignment are longer in order of group-driven, merger-driven, interaction-driven, and secularly-driven misalignment. It seems to reflect both initial position angle offsets of the galaxies and the trend between the lifetime and the physical properties.}
\end{itemize}

Given the limited computing resources, large-volume cosmological simulations usually come with low resolution in space, mass, and time.
Since the smallest (best) force resolution of Horizon-AGN is about 1\,kpc, the thin disk structure in a galaxy cannot be resolved \citep[see, ][]{2020arXiv200912373P}. 
We anticipate the next generation simulations with improved resolutions would provide more decisive information on the issue of the formation and lifetime of the star-gar misalignment.

\acknowledgments

Much of this work was conducted while SKY was visiting the University of Sydney, University of Western Australia, University of Melbourne and Swinburne University through the Astro3D Distinguished Visitor Programme from the Australian government. He thanks the generous support from the programme and the hospitality from the host institutes. SKY acknowledges support from the Korean National Research Foundation (NRF-2020R1A2C3003769). 
DJK acknowledges support from Yonsei University through Yonsei Honors Scholarship. 
JJB acknowledges support of an Australian Research Council Future Fellowship (FT180100231).
CP thanks S. Prunet for constructives comments.
Parts of this research were conducted by the Australian Research Council Centre of Excellence for All Sky Astrophysics in 3 Dimensions (ASTRO 3D), through project number CE170100013.
This work relied on the HPC resources of the Horizon Cluster hosted by Institut d’Astrophysique de Paris. We warmly thank S. Rouberol for running the cluster on which the simulation was post-processed.

%% For this sample we use BibTeX plus aasjournals.bst to generate the
%% the bibliography. The sample63.bib file was populated from ADS. To
%% get the citations to show in the compiled file do the following:
%%
%% pdflatex sample63.tex
%% bibtext sample63
%% pdflatex sample63.tex
%% pdflatex sample63.tex

%\bibliography{sample631}{}
\bibliography{ref}{}

\appendix
\restartappendixnumbering
\DeclareRobustCommand{\Xtophe}[1]{{\sethlcolor{green}\hl{#1}}}

\section{Relaxating Rings model} \label{sec:AppA}
%==============================================

Let us briefly present here a toy model that might help us understand some of the findings in the main text. In particular, we want to explain why the lifetime of misalignment should depend on the gas mass fraction. How do different physical processes affect the coupled gas and stellar disks differently? Why is cold gas fraction anti-correlated with the misalignment lifetime?

%==============================================
\subsection{Qualitative upshot} 
%==============================================

\begin{figure}
	% To include a figure from a file named example.*
	% Allowable file formats are eps or ps if compiling using latex
	% or pdf, png, jpg if compiling using pdflatex
	\includegraphics[width=\columnwidth]{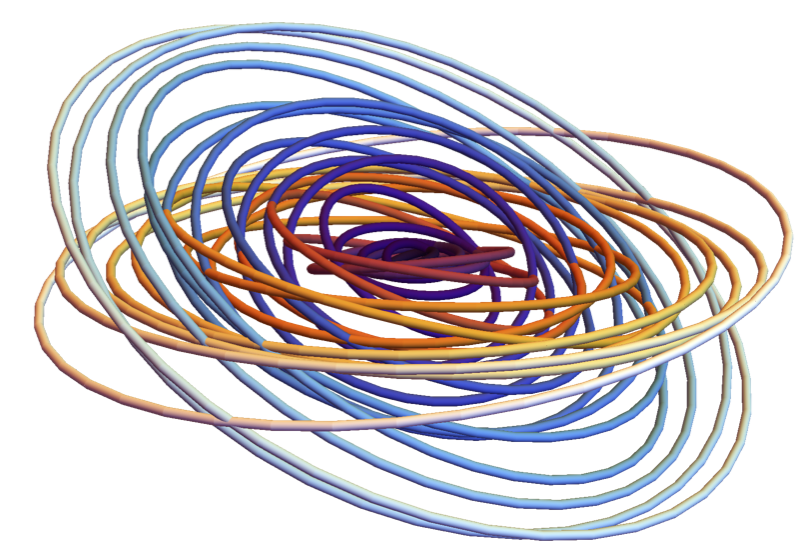}
    \caption{The relaxation of gravitationally self-interacting rings of stars and gas (in red and blue respectively). 
    The coupling can be linearized around the co-planar configuration. The equations of motions governing the N oscillators can be decoupled by moving to the eigen-frame.
}
    \label{fig:ring}
\end{figure}

The toy model describes a satellite galaxy falling into a larger group/cluster as a set of two  gravitationally-coupled rings subject to external tides.
Each set represents the gas and the stellar disk, respectively.
Each ring represents a set of orbits with a given set of actions (i.e. orbital parameters), which again for simplicity are characterized here by a mean radius and some (irrelevant) minor spread in eccentricity and vertical oscillation around circular coplanar orbits. So, in the end, the ring is completely specified by its radius and relative orientation with respect to the reference center-of-mass mid plane of the whole gas plus star system. 
Similarly, the gaseous disk is formally decomposed into such a set of coupled rings characterized by its radius $R_i$ and its orientation $(\theta_i,\phi_i)$ with respect to the mid plane.
The set of concentric gravitationally self-interacting rings is qualitatively depicted in Figure~\ref{fig:ring}. 

We will assume here that the radii do not change with cosmic time (neglecting accretion), and that the two disks are embedded into a spherical halo (or that the halo's static asphericity is accounted for in setting up the mid plane, and its time dependent variation is integrated in the `external tides' that the set of disks will be suject to).
Conversely, we will assume that the orientations of both sets of rings are time dependent, as they respond both to the tidal force imposed by the position of the satellite galaxy within the cluster, and the time dependent distortion of both (satellite and cluster) dark (sub) halos.
The gas disk is also specifically subject to two extra sets of torques induced by ram pressure and turbulent viscosity.

Let us first give a qualitative account of the expected behaviour of such a toy model before spelling out the mathematics.
Since we are concerned by departure from a set of settled co-planar disks, we will assume without loss of generality that the equations of motion describing the different rings are linearized with respect to an unperturbed co-planar configuration. 
We can therefore Taylor-expand the Hamiltonian of the system with respect to the amplitude of small oscillations of each ring above and below the (time dependent) mid plane.
In doing so, each ring will formally become a coupled oscillator with effectively one degree of freedom.
It is (tidally) coupled to all other (gas+star) rings and subject to external forcing. 
After linearization, the set of 2$N$ coupled oscillators will obey a matrix equation.
These equations of motion can be decoupled by moving to the eigen-frame diagonalizing the matrix of either gas or stellar disk.
Each eigen-mode will represents a displacement wave of the set of $N$ rings of the chosen component and will obey in isolation (neglecting briefly the other disk) a linearly forced oscillator equation, where the forcing in that frame is simply the projection of the forcing in configuration space projected onto the eigen-vectors of the matrix.
Accounting for the other disk implies that in such a frame, the evolution of the gas and star eigen-modes follows a two set of coupled forced oscillator equations as we want to explicitly account for the fact that the gaseous component can dissipate oscillations so we will add a damping term for the gas eigen-equation.

It follows quite straightforwardly that if the gas eigen-mode alone is subject to a differential torque (e.g. ram pressure) it will react to it and eventually damp it out, but it will also drag the typically more massive coupled stellar disk out of phase
so that both modes will oscillate with respect to each other for a while.
The relative amplitude of these oscillations will essentially reflect the relative mass in each disk.
Since the physical disk response is the sum of their eigen-modes, the misalignment between the disks will reflect that of their eigen-modes.

While this toy model is not meant to be taken too seriously, it has the merit of explaining qualitatively what the simulation is doing.
The key relevant ingredients are (i) the (relative) masses of the two disks i.e. the strength of their gravitational coupling; (ii) dissipation only within the gas component (iii) relative forcing on the gas component.

 \subsection{Laplace Lagrange theory of coupled damped rings}\label{sec:LLtheory}
 %==============================================

In order to be more quantitative e.g. about the relative role of gas fraction, let us spell out the toy model presented in the previous section.
This is best described in the so-called Laplace Lagrange theory.

\subsubsection{Stellar disk setup}
 %==============================================

Let us assume that the stellar orbits with guiding center $R$ in the disk are nearly coplanar ($\theta \ll 1$) and nearly circular ($e \ll 1$), where $\theta$ and $\phi$ are the polar and azimuthal angles specifying the orientation of this orbital plane.
For simplicity let us assume that we are considering the outer part of the disk, so that the potential can be described as nearly Keplerian.
Defining the canonical variables\footnote{so that $x$ and $y$ components of angular momentum obey $L_{i,x}=\gamma_i q_i$ and $L_{i,y}=\gamma_i p_i$.}, $\mathbf{p},\mathbf{q}$ as  
\begin{align}
q_i= \gamma_i \theta_i \sin (\phi_i)\,,\quad
p_i= -\gamma_i \theta_i \cos (\phi_i)\,,
\end{align} 
with $\gamma_i=\sqrt{m_i } (G M R_i)^{1/4}$, the Hamiltonian describing the coupling between the ring at radius $R_i$ in that limit reads \citep{2011MNRAS.412..187K}
 \begin{equation}
     H(\mathbf{p},\mathbf{q}) =\mathbf{p}^\mathrm{T} \cdot \mathbf{A} \cdot \mathbf{p}+
     \mathbf{q}^\mathrm{T} \cdot \mathbf{A} \cdot \mathbf{q} \,,
     \label{eq:Hamilton}
     \end{equation}
where
\begin{subequations}
  \begin{align} % requires amsmath; align* for no eq. number
     A_{ij} &=-\frac{1}{8}\frac{G m_i m_j \alpha_{ij}}{ \mathrm{max}(R_i,R_j)\gamma_i \gamma_j} b_{3/2} (\alpha_{ij})\,,\quad \mathrm{if}\quad {i\neq j}\\
    &=\sum_{k\neq i}\frac{1}{8}\frac{G m_i m_k \alpha_{ik}}{\mathrm{max}(R_i,R_k)\gamma_i^2 } b_{3/2} (\alpha_{ik})\,,\quad \mathrm{if}\quad {i= j}\,, 
\end{align} \label{eq:defA}
\end{subequations}
given
$\alpha_{ij}=
\mathrm{min}(R_i,R_j)/\mathrm{max}(R_i,R_j)$
and
\begin{subequations}
\begin{align} 
b_{3/2}(\alpha)&= \frac{2}{\pi}\int_0^\pi \frac{\cos x d x}{(1-2 \alpha \cos x +\alpha^2)^{3/2}}\,,\\
&= \frac{(1+\alpha^2)E(\alpha)-(1-\alpha^2) K(\alpha)}{\pi \alpha(1-\alpha^2)^2}\,,
\end{align} 
\end{subequations}
with $K$ and $E$ the elliptic functions of the first and second type respectively.
  
The equations of motion following from Equation~\eqref{eq:Hamilton} are best solved if we move to a frame which diagonalize the positive semi-definite symmetric matrix $\mathbf{A}$, so that in that frame, Hamilton's equation yields for each eigen-mode
\begin{equation}
\ddot {\hat q}_i + \omega_i^2 \hat q_i = \hat \xi_i\,, \label{eq:evol-equation-ring}
\end{equation}
where $\omega_i$ is the $i^\mathrm{th}$ eigen-value and $\hat \xi_i$ is the external specific force applied on the ring projected on the corresponding eigen-vector.
Here the hat quantities are (briefly) used for the new frame of coordinates.

\subsubsection{Stellar-Gas disk coupling}
%==============================================

Now let us note that so long as the surface density of the gas and the stars are proportional, the matrix $\mathbf M$ describing gas rings will be formally identical to that for the stars (up to a multiplicative factor reflecting the gas to star mass ratio), so that the eigen-space of both disks are the same.

In that frame, the gas ring eigen-mode obeys formally a similar equation to Equation~\eqref{eq:evol-equation-ring} with one extra caveat, which is that the gas can shock, so that each gas ring is subject to an extra drag force.
For expediency we consider that the drag term operates on the eigen-mode.

Finally, when considering simultaneously the evolution of both gas and star eigen-modes we need to account for their relative gravitational interaction, which can be accounted for by a supplementary coupling term in both equation.

This then leads us to consider the dynamics of the set of coupled gas plus star eigen-modes for the stars, and the gas components, whose amplitudes are written as $q_\star$, and $q_g$, respectively. 
For expediency let us consider only one such mode, (which effectively assumes that both disks can be diagonalized in the same frame, which will be the case when assuming that the stellar and gas disks are the same up to a multiplicative constant) so we drop the index $i$ following each mode and the hats.

\begin{figure}
	% To include a figure from a file named example.*
	% Allowable file formats are eps or ps if compiling using latex
	% or pdf, png, jpg if compiling using pdflatex
	\center\includegraphics[width=7.5cm]{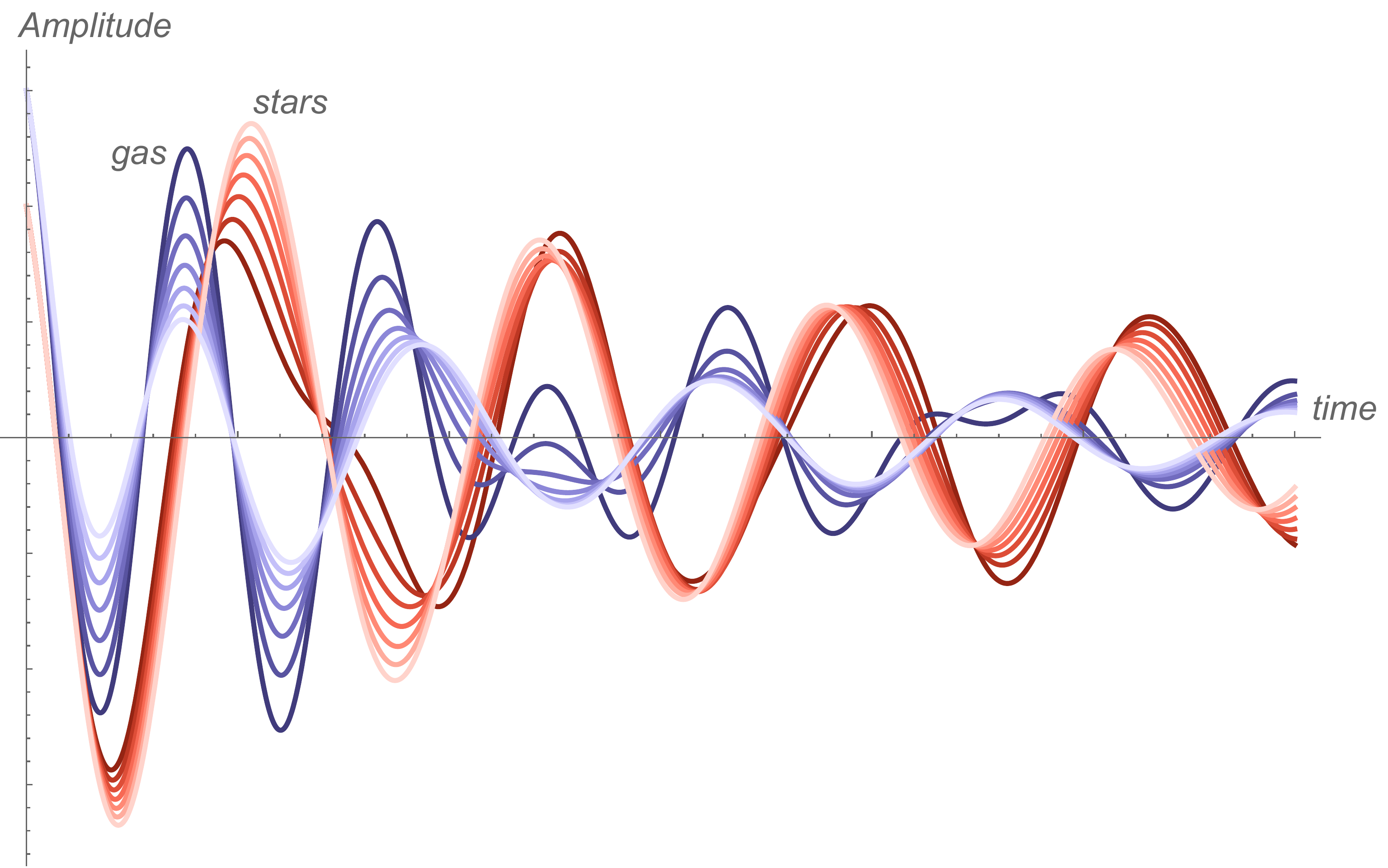}
	\center\includegraphics[width=6cm]{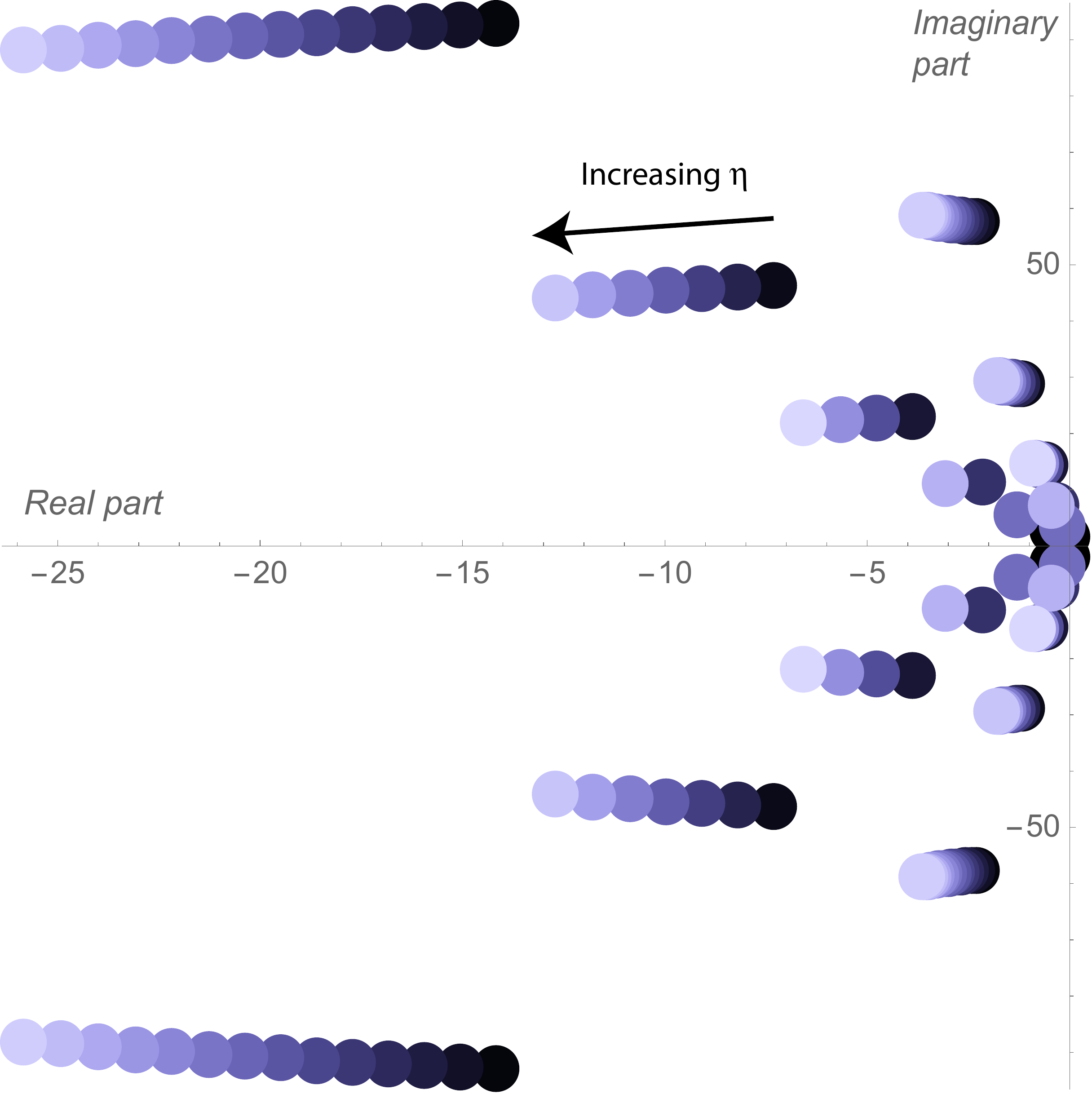}
    \caption{{\sl Top panel:} the relaxation of gravitationally self-interacting eigen-modes of gas and stars. The stars in red are driven towards the gas orientation in blue. The level of damping increase from dark to lighter curves.  
    {\sl Bottom panel:} the corresponding drift of the four roots of $S_4$, defining the frequencies of the system. Note how the real negative (damping) part increases with $\eta$, the parametrization of  dissipation within the gas disc.}
    \label{fig:realignment}
\end{figure}

We will consider that this eigen-mode has its own natural frequency, $\omega_\star$ and $\omega_g$ respectively, a coupling term, $\omega_{\star g}$ and a driving, $\xi$, and damping, $\eta$, term specific to the gas component.
The amplitude of each mode then obeys the set of coupled equations
\begin{subequations}
\begin{align} 
&\ddot { q}_\star + \omega_\star^{2} q_\star + \omega^2_{\star g} q_g \,= 0 \\
&\ddot {q}_g + \omega_g^{2} q_g + \omega_{\star g}^2 q_\star +\eta \dot { q}_g= \xi,
\label{eq:evol-equation-ring-coupled}
\end{align}
\end{subequations}
which is the main equation of this appendix. The coupling term follows from writing the Hamiltonian for the two sets of rings, while the damping and driving terms are phenomenological add on.
Note that we put all the (relative) torquing, $\xi$ on the gas component, since any torquing which applies on both components will not induce relative misalignment.
Note also that typically $\omega_\star\gg \omega_g$ given the relative mass ratio of the two disks (see, Equations~\ref{eq:defA}).
Solving for Equation~\eqref{eq:evol-equation-ring-coupled}, each stellar eigen-mode will obey
 \begin{equation}
 q_\star(t)= - \sum_{\omega \in S_4}
\frac{\omega _{g\star}^2 \displaystyle\int _{-\infty}^t \exp\left({(t-\tau ) \omega }\right) \xi\left(\tau \right)d\tau }{\eta \left(3 \omega ^2+\omega_\star^2\right)+2 \omega \left(2 \omega^2+\omega _g^2+\omega_\star^2\right)}\,, \label{eq:sol-ring-coupled}
 \end{equation}
where the frequencies, $\omega$, are one of four complex conjugate solutions of the implicit equation\footnote{note how $S_4$ reduces as it should to $\omega=\pm \omega_\star$ and $\omega=\pm \omega_g$ when both the friction and the coupling are nil.}
 \begin{equation}
S_4\!=\!\{\omega\big| \left(\omega^2+\omega_\star^2\right) \left(\omega \left(\eta+\omega \right)+\omega_g^2\right)=\omega _{g\star}^4\},
\label{eq:S4}
  \end{equation}
which will have both a damped (real) component, and an oscillatory (complex) one.

Illustratively, Figure~\ref{fig:realignment} shows the (unforced) damping of the two stellar and gas modes (of a given ad-hoc initial amplitude) when one increases the drag on the gas component (top panel), while the bottom panel shows the frequencies which are roots of $S_4$ in equation~\eqref{eq:S4}.
As expected, as one increases $\eta$, the complex roots acquire a larger and larger negative real part (corresponding to damping), and the gas disk will generally drag more efficiently the stellar disk towards itself as it settles.

%%%%%%%%%5

\begin{figure}
	% To include a figure from a file named example.*
	% Allowable file formats are eps or ps if compiling using latex
	% or pdf, png, jpg if compiling using pdflatex
	\includegraphics[width=\columnwidth]{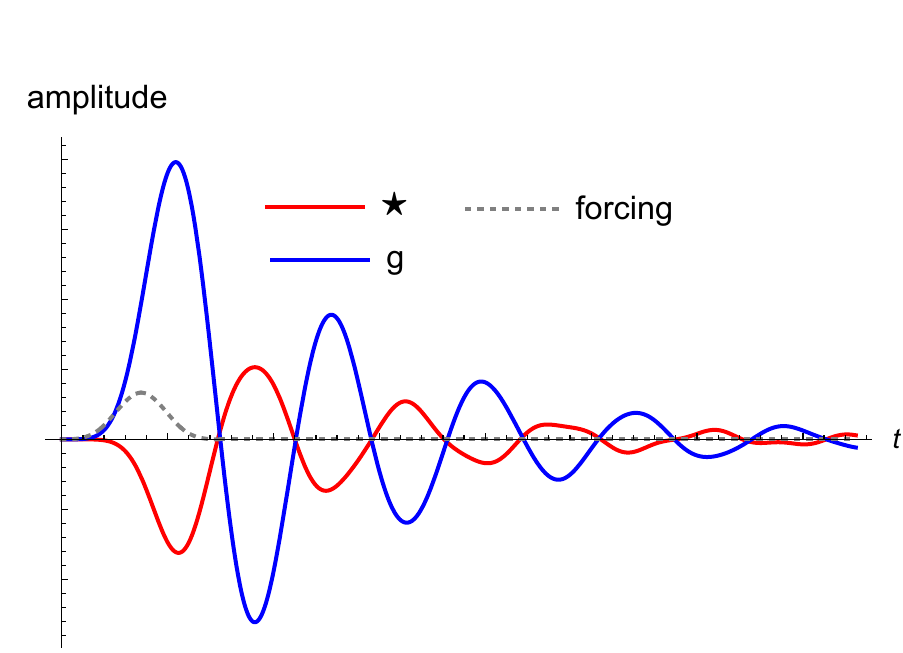}
    \caption{The lighter gas disk eigen-mode amplitude (in blue), when directly displaced by baryonic processes (e.g. ram pressure, here shown as a gray dashed impulse), will strongly become misaligned and drag the stellar disk eigen-mode (in red) first out and then back to equilibrium (since the former is subject to friction).}
    \label{fig:realignment1}
\end{figure}

%%%%%%%

\subsubsection{Stellar-Gas disk re-alignment}
%==============================================
Equipped with Equation~\eqref{eq:sol-ring-coupled}, we can now investigate the relative orientations of sets of rings corresponding to the stellar and gaseous disk respectively, to understand within the framework of the linearized Laplace-Lagrange theory how the two disks re-orient with respect to each other.
We aim here to account for the fact that only the gas disk is subject to forcing by ram pressure on the one hand, and dissipation through shocks between rings on the other hand. 
The gas disk is also typically lighter, hence more responsive to torques.

Let us first consider an idealized experiment when only the gas disk is subject to an impulse which propagates to the other disk before eventually both modes damp and the coupled system settles.  
Figure~\ref{fig:realignment1} shows qualitatively the result of such experiment. 
As expected the gas response is significantly stronger because the gas disk is assumed to be here less massive, in agreement with the findings of the main text. 
It reacts in phase to the impulse, while the stars are out of phase. 
As shocks and turbulence in the gas component dissipate energy, both oscillations eventually de-phase and damp away.

\begin{figure}
	% To include a figure from a file named example.*
	% Allowable file formats are eps or ps if compiling using latex
	% or pdf, png, jpg if compiling using pdflatex
	\includegraphics[width=\columnwidth]{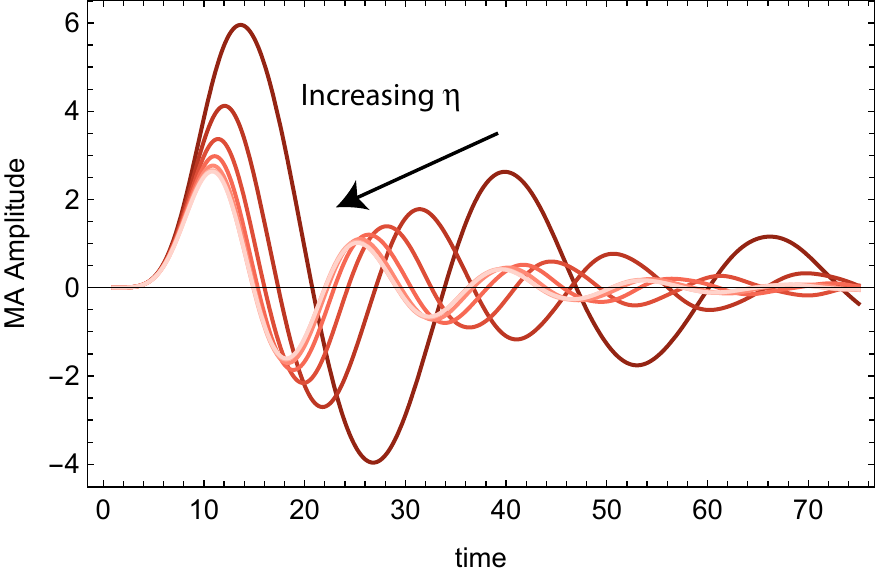}
    \caption{The relative misalignment amplitude of gravitationally self-interacting eigen-modes of gas and stars which are subject to a given initial displacement. 
    The level of damping and the mass of the gas disk increase from dark to lighter curves. 
    The net effect is that the settling takes longer for lighter disks as can be seen from the envelope of the dark red curves.
    The lifetime of misalignment anti-correlates with cold gas fraction.}
    \label{fig:realignment2}
\end{figure}

Focusing now specifically on the quantities of interest in our paper, Figure~\ref{fig:realignment2} illustrates the damping of two modes when one increases both the drag on the gas disk and its mass. 
As expected, the lighter the gas disk, the longer the settling phase.
This is also in qualitative agreement with the findings of the main text, corresponding to the situation where a given galaxy enters a group or a cluster and the gas component feels ram pressure from the hot corona.
As expected from the toy model and observed in the simulation, the more gas rich the disk is, the stronger the turbulence, the faster the damping the shorter the lifetime of misalignment.

%==============================================
\subsection{Discussion} \label{sec:conclusion}
%==============================================

As previously stated, the simplistic ring model is only illustrative and not without flaws. 
Within the context of this paper, we do not aim to quantify the statistical properties of the forcing, which one could extract from the sets of measured misalignment events.
It would also be of interest, beyond the scope of this appendix, to quantify the damping process.

Note that to be more quantitative the misalignment should be measured in configuration space which can be extracted from the eigen-space information.
Note that non-linear mode coupling could and will also dampen oscillations, as will angular momentum exchanges at resonances within the stellar disk.

At some very coarse level, one can consider that fly-bys correspond to an extreme case of 
satellite-cluster interaction, so the arguments presented for the latter would apply to the former.
Similarly, cosmic infall could be approximated by a secularly added misaligned ring which will inject some energy into the set of coupled oscillators.

\bibliographystyle{aasjournal}

%% This command is needed to show the entire author+affiliation list when
%% the collaboration and author truncation commands are used.  It has to
%% go at the end of the manuscript.
%\allauthors

%% Include this line if you are using the \added, \replaced, \deleted
%% commands to see a summary list of all changes at the end of the article.
%\listofchanges

\end{document}